\documentclass[%
 aip,
 jmp,%
 amsmath,amssymb,
 reprint,%
 nofootinbib,%
 floatfix, %
]{revtex4-1}

\usepackage{graphicx}
\usepackage{caption}
\usepackage{subcaption}
\usepackage{dcolumn}
\usepackage{bm}
\usepackage[dvipsnames]{xcolor}
\usepackage{amssymb}
\usepackage{verbatim}
\usepackage{appendix}
\usepackage{listings}
\usepackage{units}
\usepackage{fancyhdr}
\usepackage{csquotes}
\usepackage{hyperref}

\usepackage{etoolbox}
\makeatletter
\patchcmd{\hyper@makecurrent}{%
    \ifx\Hy@param\Hy@chapterstring
        \let\Hy@param\Hy@chapapp
    \fi
}{%
    \iftoggle{inappendix}{
        \@checkappendixparam{chapter}%
        \@checkappendixparam{section}%
        \@checkappendixparam{subsection}%
        \@checkappendixparam{subsubsection}%
        \@checkappendixparam{paragraph}%
        \@checkappendixparam{subparagraph}%
    }{}%
}{}{\errmessage{failed to patch}}

\newcommand*{\@checkappendixparam}[1]{%
    \def\@checkappendixparamtmp{#1}%
    \ifx\Hy@param\@checkappendixparamtmp
        \let\Hy@param\Hy@appendixstring
    \fi
}
\makeatletter

\newtoggle{inappendix}
\togglefalse{inappendix}

\apptocmd{\appendix}{\toggletrue{inappendix}}{}{\errmessage{failed to patch}}
\apptocmd{\subappendices}{\toggletrue{inappendix}}{}{\errmessage{failed to patch}}

\newcommand*\obar[2][0.75]{
    \sbox{\myboxA}{$\m@th#2$}%
    \setbox\myboxB\null
    \ht\myboxB=\ht\myboxA%
    \dp\myboxB=\dp\myboxA%
    \wd\myboxB=#1\wd\myboxA
    \sbox\myboxB{$\m@th\overline{\copy\myboxB}$}
    \setlength\mylenA{\the\wd\myboxA}
    \addtolength\mylenA{-\the\wd\myboxB}%
    \ifdim\wd\myboxB<\wd\myboxA%
       \rlap{\hskip 0.5\mylenA\usebox\myboxB}{\usebox\myboxA}%
    \else
        \hskip -0.5\mylenA\rlap{\usebox\myboxA}{\hskip 0.5\mylenA\usebox\myboxB}%
    \fi}
\makeatother

\def\smalloint{\begingroup\textstyle \oint\endgroup}

\newcommand{\ordo}[1]{{\cal O}\left( #1 \right)}

\renewcommand{\vec}[1]{\boldsymbol{#1}}

\newcommand{\e}{\ensuremath{\mathrm{e}}}
\newcommand{\p}{\ensuremath{\partial}}

\renewcommand{\d}{\ensuremath{\mathrm{d}}}

\newcommand{\remove}[1]{{}}

\newcommand{\psio}{\ensuremath{\psi^o}}
\newcommand{\perfect}{\textsc{Perfect}}
\newcommand{\DKE}{drift-kinetic equation}
\newcommand*{\nouncite}[1]{Ref.~\citenum{#1}}

\newcommand{\Dbulk}{blue}
\newcommand{\Hebulk}{red}

\newcommand{\appref}[1]{\hyperref[#1]{Appendix~\ref{#1}}}

\usepackage{morefloats}
\begin{document}

\preprint{AIP/123-QED}

\title{Neoclassical flows in deuterium-helium plasma density pedestals}
\author{S. Buller}
\email{bstefan@chalmers.se}
\affiliation{Department of Physics, Chalmers University of Technology,
  SE-41296 G\"{o}teborg, Sweden}
\author{I. Pusztai}
\affiliation{Department of Physics, Chalmers University of Technology,
  SE-41296 G\"{o}teborg, Sweden}
\author{S.L. Newton}
\affiliation{Department of Physics, Chalmers University of Technology,
  SE-41296 G\"{o}teborg, Sweden}
\affiliation{CCFE, Culham Science Centre, Abingdon, Oxon OX14 3DB, UK}
\author{J.T. Omotani}
\affiliation{Department of Physics, Chalmers University of Technology,
  SE-41296 G\"{o}teborg, Sweden}




\date{\today}

\begin{abstract}
  In tokamak transport barriers, the radial scale of profile
  variations can be comparable to a typical ion orbit width, which
  makes the coupling of the distribution function across flux surfaces
  important in the collisional dynamics.  We use the radially global
  steady-state neoclassical $\delta f$ code \perfect{} to calculate
  poloidal and toroidal flows, and radial fluxes, in the pedestal.  In
  particular, we have studied the changes in these quantities as the
  plasma composition is changed from a deuterium bulk species with a
  helium impurity to a helium bulk with a deuterium impurity, under
  specific profile similarity assumptions. The poloidally resolved
  radial fluxes are not divergence-free in isolation in the presence
  of sharp radial profile variations, which leads to the appearance of
  poloidal return-flows. These flows exhibit a complex radial-poloidal
  structure that extends several orbit widths into the core and is
  sensitive to abrupt radial changes in the ion temperature gradient.
  We find that a sizable neoclassical toroidal angular momentum
  transport can arise in the radially global theory, in contrast to
  the local.
\end{abstract}

\keywords{neoclassical transport, tokamak, radially global, pedestal, flows, deuterium, helium}
\maketitle

\section{Introduction}
Transport barriers are regions in a magnetically confined plasma with
reduced cross-field turbulent transport, which leads to steep
gradients in density and temperature. Such formations have a major
impact on fusion plasmas, as they can enable fusion relevant
conditions to be reached in smaller devices. Thus, practically all
plans for future magnetic fusion experiments and reactors include at
least an edge transport barrier, commonly known as \emph{the
  pedestal}.

Since the discovery of transport barriers (the ELMy H-mode
\cite{pedestalASDEX}) there have been numerous studies
\cite{biglari1990shearedVp,Staebler1991,LaHaye1995VHflowShear,burrell1997,Highcock2011}
indicating the crucial role plasma flows play in the transition to
improved confinement regimes.  In particular, the equilibrium flows
have been experimentally observed to play a major role in setting the
transition threshold\cite{ASDEX_L-H_Er}.  Thus well resolved plasma
flow measurements in the pedestal, especially those of the main ion
species, are highly desirable for progress in the understanding of the
barrier formation.

However, the flows of fusion relevant hydrogen ions are challenging to
infer directly due to their involved emission spectrum, which
complicates the use of standard charge exchange techniques (see
\nouncite{haskey2016flowDIIID} for a recent effort to tackle these
difficulties in DIII-D).  Instead, flow diagnostics often rely on
measuring the flows of some impurity species, such as
He\cite{silburn2014}, B\cite{churchill2015PolAsymETB} and
C\cite{silburn2014,menmuirJETcxrs2014}, from which the deuterium flows
are inferred. Yet another technique is to measure the main ion flows
in non-hydrogenic plasmas; this approach has been used with helium
plasmas, in DIII-D\cite{DIIIDHePlasma1994}, and recently in ASDEX
Upgrade\cite{wolfrumASDEXped2015}.

That being said, theoretical calculations are required to relate these
flows to deuterium flows. Specifically, it is important to know how
impurity ion flows relate to main ion flows, and how the flows in
helium and deuterium bulk plasmas compare to each other. These
questions have previously been addressed with local neoclassical
predictions\cite{kimDiamond1991,testa2002}.  Such local
calculations assume a scale separation between the radial variations
in plasma profiles and the orbit width, thus they are not necessarily
applicable to the H-mode pedestal, where profiles can vary
significantly over an orbit width.  It is in the pedestal where flow
measurements would be particularly interesting in order to study the
H-mode, and neoclassical effects may be expected to contribute most
strongly as a result of the reduced turbulent transport.

To address these questions in relation to the pedestal, we numerically
investigate neoclassical flows in deuterium and helium plasmas, using
the neoclassical global $\delta f$ code \perfect{} (Pedestal and Edge
Radially-global Fokker-Planck Evaluation of Collisional
Transport)\cite{landreman12,landreman2014}. The model profiles used
here follow the approach of \nouncite{pusztai2016perfect} to capture
features of experimental JET pedestals\cite{ELMyHModeMaggi2015} in
multi-species simulations, while remaining within the validity of the
$\delta f$ theory.

Due to the radial coupling in the global theory, the resulting fluxes
are not divergence free within flux-surfaces. Thus flows within flux
surfaces should be considered together with the radial flows. This has
been considered as an explanation for poloidal impurity flow
observations in
\nouncite{putterichASDEXpoloidalAsymParallelRotation2012}, although
modeling of the radial-poloidal flow structures was outside the scope
of their study.  In this paper, we discuss radial transport alongside
poloidal and toroidal flows.

In \autoref{sec:meth}, we describe the radially
global $\delta f$ theory, including constraints posed by our
linearization about a Maxwell-Boltzmann distribution.
In \autoref{sec:re}, we relate the assumptions presented in the
previous section to design choices in our numerical study, and
present two bulk helium scenarios based on different
similarity assumptions. Flows in both local and global simulations are
presented, and we find that the global poloidal flows display notably
larger poloidal and radial variations, with odd-parity in-out
structure, which is sensitive to changes in the temperature
gradient. Poloidally resolved radial flows are then
considered together with poloidal flows. Finally, we consider the
total radial transport of particles, heat and toroidal angular momentum. In
global simulations we observe order unity modifications to the local particle
and heat fluxes, both inside the pedestal and a few orbit widths into the
core. More importantly the angular momentum flux can
become significant in the global simulations.

\section{Global $\delta f$ drift-kinetics}
\label{sec:meth}
Assuming that there is a sufficient scale separation between
  the gyroradius and the background profiles, and excluding fast
  transients, collisional transport can
  be calculated from the drift-kinetic equation (DKE)
\begin{equation}
  \frac{\p f}{\p t} + \left( v_\| \vec{b} + \vec{v}_d \right) \cdot
  \nabla f = C[f].\label{dke}
\end{equation}
The terms in the above equation are: $f$, the gyroaveraged
distribution function; $v_\| = \vec{v}\cdot\vec{b}$, with $\vec{v}$
the velocity, $\vec{b} \equiv \vec{B}/B$, with $\vec{B}$ the magnetic
field and $B\equiv |\vec{B}|$; the drift velocity $\vec{v}_d$ -- which
we decompose into $E\times B$ and magnetic drifts $\vec{v}_{d} =
\vec{v}_E + \vec{v}_{m}$, $\vec{v}_{m} = v_\|^2 \Omega^{-1} \nabla
\times \vec{b} +v_\perp^2 (2\Omega B^2)^{-1} \vec{B} \times \nabla B$,
$\vec{v}_E = B^{-2} \vec{E} \times \vec{B}$; $\Omega = ZeB/m$ is the
gyrofrequency, with $e$ the elementary charge, $Ze$ the species
charge; $\vec{v}_\perp = \vec{v} - v_\| \vec{b}$ is the velocity perpendicular to $\vec{b}$. $C$ is a collision
operator -- here the Fokker-Planck operator.  
The gradients are taken with total energy $W = mv^2/2 + Ze\Phi$ and magnetic moment $\mu
= mv_\perp^2/(2B)$ held constant, where $\Phi$ is the electrostatic
potential.

To obtain a linear theory, we expand the distribution function $f$
around a flux function, stationary Maxwell-Boltzmann distribution
$f_M(\psi)$, $f=f_M + f_1+\mathcal{O}(\delta^2 f_M)$, $f_1/f_M =
\ordo{\delta}$. Here $2\pi\psi$ is the poloidal magnetic flux, which
we use as our flux-surface label; $\delta$ is an expansion parameter
representing the smallness of the thermal ion orbit width compared to
the scale of the device, to be defined more rigorously once we derive
further validity conditions for the linearization.  In general, a
physical quantity $X$ is decomposed as $X = X_0 + X_1$. From the
decomposition of the distribution function, it follows that the zeroth
order density $n_0$ and temperature $T_0$ profiles are flux functions;
and that the total flow velocities in the laboratory frame are much
smaller than the thermal speed of the species (i.e. ``low-flow
ordering'').  In addition, we assume $\Phi_0 = \Phi_0(\psi)$. These
assumptions are made for the sake of convenience, and do not represent
an inherent limitation of the $\delta f$ framework\footnote{One
  possible generalization is to include order unity density variations
  on a flux surface, which naturally develop for impurities in the
  pedestal\cite{fulop99,fulop01,landreman11,churchill15,pusztaiRFNEO}. Such
  poloidal variations have been implemented in the stellarator code
  {\sc Sfincs}\cite{SFNICSmanual,sfincs2014} that shares a very similar
  numerical framework to {\sc Perfect}.}.

Defining the non-adiabatic part of $f_1$ as
\begin{equation}
  g = f_1 -\frac{Ze\Phi_1}{T} f_M,\label{g}
\end{equation}
we linearize \eqref{dke} to obtain
\begin{equation}
  \left( v_\| \vec{b} + \vec{v}_{d0} \right) \cdot (\nabla g)_{W_0,\mu}
  -C_l[g]  = - \vec{v}_m \cdot (\nabla f_M)_{W_0,\mu},\label{ldke}
\end{equation}
where $C_l$ is the linearized Fokker-Planck operator, $W_0 = mv^2/2 +
Ze\Phi_0$; $\Phi_1$ has been eliminated from the drift
velocity $\vec{v}_{d0} = B^{-2} \vec{B} \times \nabla \Phi_0 + \vec{v}_{m}$
and we have assumed a steady state distribution, $\p f/\p t = 0$.

We are now in a position to more precisely define our expansion
parameter $\delta$.
By construction, $g/f_M = \ordo{\delta}$, thus $\delta$ measures the size of $g$. The size of $g$ scales with
the inhomogeneous term on the right-hand side of \eqref{ldke}, which can
be approximated as\cite{landreman2014}
\begin{equation}
  \vec{v}_m \cdot (\nabla f_M)_{W_0,\mu}= \vec{v}_m \cdot \nabla \psi
  \left.\frac{\p}{\p \psi}\right|_{W_0} \!\left[ \left(\tfrac{m}{2\pi
      T_0}\right)^{3/2} \eta_0 e^{\frac{W_0}{T_0}}\right],
\label{vmdotgrad}
\end{equation}
where $W_0$ is a constant
with respect to our derivative, and we have introduced the \emph{pseudo-density}
\begin{equation}
  \eta_0(\psi)= n_0 e^{Ze\Phi_0/T_0}.\label{eta}
\end{equation}
(Henceforth, to streamline notation, the subscript ``$0$'' will be
dropped.)  It follows from
\eqref{vmdotgrad} that only the temperature and $\eta$ gradients enter
into the inhomogeneous term in \eqref{ldke} and thus set the size of
$g$ and act to make the distribution function
non-Maxwellian. To characterize the size of the gradients of a plasma parameter $X$ we introduce
\begin{equation}
  \delta_X \equiv \frac{\rho_p}{L_X},
\label{deltaX}
\end{equation}
where $\rho_p = m v_T/(Ze B_p)$ is the poloidal gyroradius, with the
poloidal magnetic field $B_p$, and thermal velocity $v_T = \sqrt{2T/ m}$;
and $L_X = -[d (\ln X)/dr]^{-1}$ is the scale length of
  $X$. Since the size of the departure from $f_M$ depends on the
  gradients in $\eta$ and $T$, we can finally give a conservative
  definition of our expansion parameter $\delta$ as 
\begin{equation}
\delta=\max_{\psi;\,X=\eta_a,T_a}  \delta_X(\psi) \ll 1,
\label{delta}
\end{equation}
where $a$ is a species index. In words, $\delta$ is the largest value of
the $\delta_X$ profiles corresponding to $T_a(\psi)$ and
$\eta_a(\psi)$ for all species.

Note that while $n$ and $\Phi$ can vary on shorter length scales if the corresponding $\eta$ has a slow radial variation, the DKE \eqref{dke} itself is based on an expansion in $\rho/L_X = (B_p/B) \delta$ for all plasma parameters,
so no plasma profile may have gyroradius scale variation. We also require $B_p/B \ll 1$ to ensure that the gyroradius and the orbit-width expansions can be performed separately.

Equation \eqref{ldke} is the global linearized drift-kinetic
  equation, from which the conventional local $\delta f$
  equation can be recovered by dropping the $\vec{v}_{d0} \cdot \nabla g$ term.
  The latter is solved when we refer to ``local'' solutions.
  This corresponds to ordering $\nabla g = \ordo{\delta f_M/L}$, meaning that the radial variation of $g$ -- and thus the radial variation of radial fluxes\cite{landreman12} -- appear at order $(B_p/B) \delta^2 n v_T/L$ in \eqref{ldke}.
Note that neglecticing $\vec{v}_m \cdot \nabla g$ means that $\psi$ only appears as a parameter in the local equation, while the global equation is a differential equation in $\psi$, and thus requires radial boundary conditions. (In this work, we use solutions to the local equation as boundary conditions; see the first part of \autoref{sec:re} for details.)
 
In the pedestal region, $\vec{v}_{d0} \cdot \nabla g$ must be retained, as the \emph{radial variation} of the fluxes is in general not small in the presence of sharp background profiles.
However, radially varying steady-state particle fluxes are inconsistent with particle conservation, unless they are compensated for by sources in the DKE \eqref{ldke}\cite{landreman12}.
  
  These sources may represent real sources, due to atomic physics processes,  but could also include the radial
  variation of other fluxes -- due to e.g.~turbulence -- which are excluded
  from our modeling but needed to cancel the radial variation of the modelled fluxes. We thus add a source term $S$ to \eqref{ldke} and obtain
\begin{equation}
  \left( v_\| \vec{b} + \vec{v}_{d0} \right) \cdot (\nabla g)_{W,\mu}
  - C_l[g] = - \vec{v}_m \cdot (\nabla f_M)_{W,\mu}+S,\label{ldkes}
\end{equation}
which is the equation we solve when we refer to ``global'' solutions.

The velocity and poloidal dependence of the sources are specified to
yield particle and heat sources, while the radial dependence of these
sources is solved for alongside $g$ by imposing that the perturbed
distribution function should have zero flux-surface averaged density
and pressure. This means that the input profiles specify the flux
surface averages, that is
  \begin{align}
    \textstyle X_0 &= \langle X \rangle \equiv (V')^{-1}\smalloint
    X(\vec{B}\cdot \nabla \theta)^{-1}\d\theta, 
    \\ \textstyle V'
    &\equiv \smalloint (\vec{B}\cdot \nabla\theta)^{-1} \d\theta,
    \end{align}
where $\theta$ is an angle-like poloidal coordinate, and we
  introduced the flux surface average, denoted by $\langle\cdot\rangle$.
  In this work, we assume that the sources have no poloidal dependence,
  and have the same velocity structure as in \nouncite{landreman2014}.

\section{Simulations and results}
\label{sec:re}
\label{ssec:He}
We consider a plasma with deuterium as bulk species and helium as
impurity, and a bulk helium plasma with a deuterium impurity species,
as two extreme cases of an ion concentration scan.

We would like to focus on differences in neoclassical phenomena that
are linked to the charge and mass of the various ion components; our
philosophy is thus to minimize changes in the plasma profiles as the
ion composition is changed. In reality, plasma with different ion
composition can have significantly diffrent profiles, for reasons
ranging from basic physical differences related to mass and charge
dependences of various phenomena\cite{pusztai2011isotope} to more
practical ones, such as differences in heating or
recycling\cite{bessenrodtWeberplas1993isotopeASDEX}.

Forgoing these complications, we consider two scenarios, with profiles based on different
similarity assumptions, which we refer to as \emph{fixed $\Phi$} and \emph{fixed $n_e$}.
The different names indicate which profile in the quasi-neutrality
condition,
\begin{equation}
n^{}_{\text{e}} = Z^{}_\text{D}\eta^{}_{\text{D}} e^{-Z^{}_{\text{D}}e\Phi/T^{}_{\text{D}}} + Z^{}_\text{He}\eta^{}_{\text{He}} e^{-Z^{}_{\text{He}}e\Phi/T^{}_{\text{He}}},
\label{quasineu}
\end{equation}
is kept fixed when changing from a D to He bulk plasma.
It is convenient to express this relation in terms of the ion
pseudo-densities and temperatures, as these are constrained by \eqref{delta}.

For the ion pseudo-density profiles we use linear profiles, with a
slope based on the experimental JET outer core $n_e$ profile of Figure
16 in \nouncite{ELMyHModeMaggi2015}; these core gradients generally
satisfy the constraints posed by \eqref{delta}.  Likewise, our ion
temperature profiles were based on the core temperature gradient of
the same figure. In addition, we further reduce the ion temperature
gradients in the core to make a proxy for the local deuterium heat
flux, $Q^{}_\text{D} \sim n^{}_\text{D} T_{\text{D}}^{3/2} \p_\psi
T^{}_{\text{D}}$, equal at the boundaries. This change in gradients
both reduces the total heat and particle sources, when integrated over
the simulation domain, while it also leads to a localization of
sources around the pedestal region\cite{landreman2014}. We will
discuss the effect of changes in the temperature gradient in
\autoref{sec:flowCPhi}.

The electron temperature profile is allowed to have structure on the
ion orbit width scale, and thus uses the full temperature profile in
Figure 16 of \nouncite{ELMyHModeMaggi2015}. We note that, as a
consequence, the electron flows are much larger than the ion flows in
absolute magnitude, and thus the bootstrap current is dominated by the
electron contribution. As a result, the fixed $n_e$ similarity class
keeps the bootstrap current profile practically fixed.

In the fixed $\Phi$ scenario, $\Phi$ is then chosen to make $n_{\rm
  D}$ similar to the experimental $n_{\rm e}$ profile. Finally,
$n_{\rm He}$ is determined by $\Phi$, $T^{}_{\text{He}}$ and
$\eta^{}_{\rm He}$.  Bulk D and He plasmas are obtained by rescaling
the ion $\eta$ profiles relative to each other, which yields electron
density profiles that vary with the ion concentrations -- while
leaving $\delta_\eta$ unaffected.  Although the shape of $n_e$ varies,
the ion profiles are re-scaled to yield the same electron density at
the pedestal top.

In the fixed $n_e$ scenario, $n_e$ is instead specified from the experimental $n_e$ profile,
and $\Phi$ is obtained from the quasi-neutrality
condition \eqref{quasineu}, and thus depends on the helium
concentration. Note that for a pure deuterium plasma, the methods
yield the same $\Phi$, and thus the bulk deuterium simulations are
approximately the same in both similarity classes.

The pedestal usually extends
outside the last closed flux surface (LCFS), while our model is restricted to
closed field lines. We consider the region which would be in the open
field line region as a numerical buffer region of the simulation.
Such a region is introduced to better accommodate the outer radial boundary
condition imposed in the simulation: a solution to the \emph{local} \DKE{}
is imposed as boundary condition where particles enter the radial
domain.
For the local \DKE{} to be valid, the density profiles in the buffer region have artificially reduced, core relevant, gradients. Since the results in the buffer region are not necessarily physically meaningful, we do not show this region in the figures.

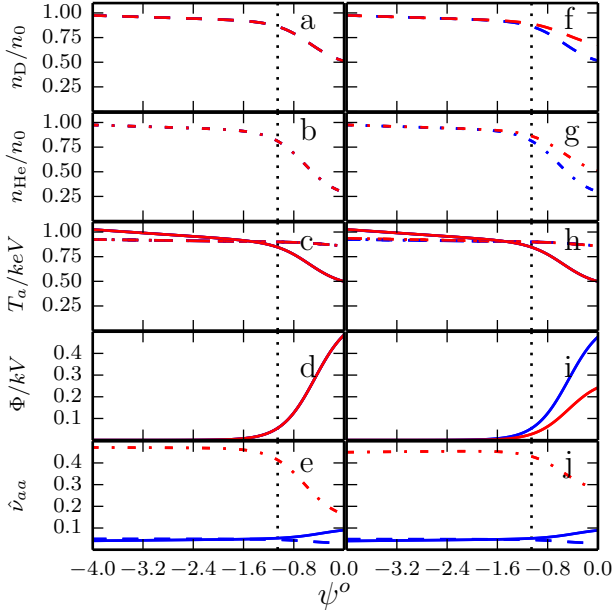
\begin{figure}
  \includegraphics{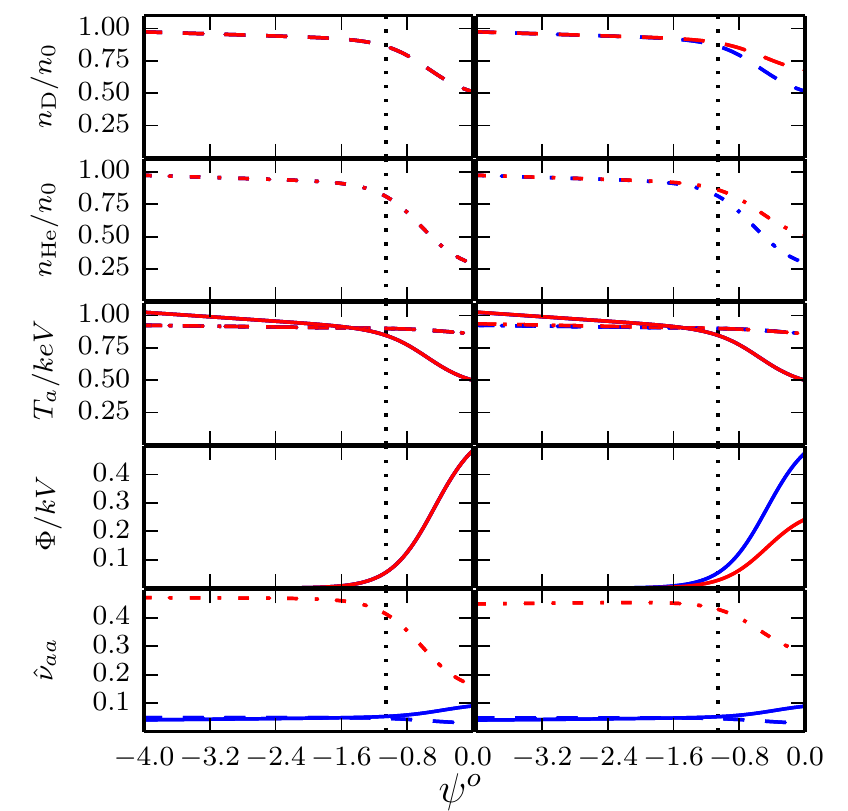}
\put(-125,220){\large a}
\put(-125,178){\large b}
\put(-125,136){\large c}
\put(-125,88){\large d}
\put(-125,52){\large e}
\put(-25,220){\large f}
\put(-25,178){\large g}
\put(-25,136){\large h}
\put(-25,88){\large i}
\put(-25,52){\large j}
  \caption{\label{fig:dHeinputs}Input profiles (a-d,f-i) used in the
  fixed $\Phi$ scan (left column) and fixed $n_e$ scan (right
  column), and the collisionality (e,j) for the different species in
  the simulations in which their density is highest. Solid: electrons,
  dashed: deuterium, dash-dotted: helium. Bulk species is deuterium
  (\Dbulk{} curves), or helium (\Hebulk{} curves), note that the curves overlap.}
\end{figure}

The resulting input profiles are depicted in \autoref{fig:dHeinputs},
for the fixed $\Phi$ and fixed $n_e$ scan (left and right columns,
respectively). In both scans the \Dbulk{} profiles are for
bulk-deuterium plasmas ($n_\text{He}/n_\text{D} =0.01$ in the core region)
and \Hebulk{} lines are for bulk-helium plasmas
($n_\text{D}/n_\text{He} =0.01$); solid lines are electron profiles,
dashed lines are deuterium and dash-dotted lines helium. As a radial
coordinate we use
\begin{equation}
\psio = \frac{\psi - \psi^{}_{\text{LCFS}}}{ \left\langle\frac{1}{B_p}\frac{\p \psi}{\p r}\right\rangle \sqrt{\epsilon T^{}_\text{D} m^{}_\text{D}}/e},
\end{equation}
which is the poloidal flux normalized and shifted so that $\psio=0$ at
the LCFS and a unit change of $\psio$ corresponds to a typical trapped
thermal ion orbit width at $\psi$; $\epsilon$ is the inverse aspect
ratio (defined as in Miller geometry\cite{miller98}).  The last
subfigures in \autoref{fig:dHeinputs} depict the species
self-collisionalities
\begin{equation}
\hat{\nu}_{aa} = \frac{\nu_{aa}}{v_T/qR_0} = \frac{qR_0}{12 \pi^{3/2}}  
\frac{e_a^4n_a\ln\Lambda}{\epsilon_0^2T_a^{2}},
\label{nuhat}
\end{equation}
in the simulations that yield the highest collisionality for the
specific species: $\hat{\nu}^{}_\text{DD}$ and
$\hat{\nu}^{}_\text{ee}$ are given for the
bulk D plasma, and $\hat{\nu}^{}_\text{HeHe}$ is given for bulk He.  Here, $q$ is the safety
factor, $R_0$ the major radius at the magnetic axis, and $\nu_{aa}$
the self-collision frequency of species $a$\cite{helander2005}.  Since $\hat{\nu}_{aa} \ll 1$,
all our simulations have all species in the banana regime
-- with the bulk helium plasma being an
exception, where $\hat{\nu}_\text{HeHe} \approx 0.5$.

In the following figures, curves are color coded as in as in
\autoref{fig:dHeinputs} to indicate ion composition. In addition,
dashed lines indicate output from local simulations, and solid lines
are global results. The same color and line styles are applied to the
frames of 2D plots.  The following normalization is used throughout
the paper. Quantities with a hat are normalized to a reference
quantity that is species-independent in most cases, $\hat{X} =
X/\bar{X}$.  The reference quantities used in this work are:
$\bar{R}=\unit[3.8]{m}$, $\bar{B}=\unit[2.9]{T}$,
$\bar{n}=\unit[10^{20}]{m^{-3}}$, $\bar{T}=e\bar{\Phi}=\unit[1]{keV}$,
$\bar{m}=m^{}_\text{D}$, where $m^{}_\text{D}$ is the deuterium
mass. These numbers are based on ``typical'' quantities for JET, and
only affect the normalization of the results.  From these, we define a
reference speed as $\bar{v}=\sqrt{2\bar{T}/\bar{m}}$, and the
dimensionless constant $\Delta=
\bar{m}\bar{v}/(\bar{e}\bar{B}\bar{R})$, which corresponds to a
normalized gyroradius at the reference quantities.  Specifically, we
have: fluid flow velocity $\hat{\vec{V}} = \vec{V} /(\Delta \bar{v})$;
sources $\hat{S}_a = \bar{v}^2\bar{R}
S_a/(\Delta\bar{n}\hat{m}_a^{3/2})$; flux surface incremental volume
$\hat{V}' = (\bar{B}/\bar{R}) d_\psi V $, particle flux,
$\hat{\vec{\Gamma}}_a = \int d^3v g_a \vec{v}_{ma}/(\bar{n}\bar{v})$;
toroidal momentum flux (divided by mass), $\hat{\vec{\Pi}}_a = \int
d^3v g_a v_\parallel I \vec{v}_{ma}/(\bar{n}\bar{v}^2\bar{R} B)$, with
$I(\psi)=RB_t$, $R$ the major radius and $B_t$ the toroidal magnetic
field; heat flux, $\hat{\vec{Q}}_a =\int d^3v g_a m_a
v^2\vec{v}_{ma}/(2 \bar{T}\bar{n}\bar{v})$; conductive heat flux,
$\hat{\vec{q}}_a =\hat{\vec{Q}}_a - (5/2)\hat{T}\hat{\vec{\Gamma}}_a$.
In addition, we define the normalized scalar radial particle flux
\begin{equation}
\hat{\Gamma}_a = \frac{\hat{V}'}{\Delta^{2}\pi\bar{R}\bar{B}} 
  \langle \hat{\vec{\Gamma}}_a \cdot \nabla \psi\rangle, 
\end{equation}
and analogously, scalar fluxes for heat and momentum. When comparing poloidal and radial fluxes (as in \autoref{fig:stream}), it is convenient to instead project the radial fluxes on a unit vector, in which case we use $\vec{\Gamma}_a \cdot \vec{\psi}$, with $\vec{\psi} = \nabla \psi/(R B_p)$ being the unit vector in the $\nabla\psi$-direction.


\subsection{\label{ssec:flow}Flows}
First we study the radial and poloidal structure of the particle
flows. As we will see, the flows in global simulations exhibit
qualitatively different features from the local results. In local
theory the flows should be divergence-free on each flux surface in
isolation. This is broken in the global case where radial flows play
an important role in making the total flow divergence-free.  Before
discussing the flows in the fixed $\Phi$ and fixed $n_e$ similarity
classes, we express the toroidal and poloidal flows, $V_t$ and $V_p$,
in terms of $g$,
\begin{align} \begin{aligned} 
V_p =& \frac{B_p}{nB} \int d^3 v
  v_\| g + \frac{T}{mB\Omega} IB_p \left[\frac{p'}{p}
  + \frac{Ze\Phi'}{T}\right] \\ +&\frac{IB_p}{nB^2} \Phi' \int
  d^3 v g + \frac{IB_p}{2nB\Omega} \frac{\p}{\p\psi} \int d^3 v
  v_\perp^2
  g \end{aligned}
\raisetag{3\baselineskip}\label{eq:Vp}\\ 
\begin{aligned}
  V_t =& \frac{B_t}{nB} \int d^3 v v_\| g -\frac{T}{mB\Omega}
  B_p^2R \left[\frac{p'}{p} + \frac{Ze\Phi'}{T}\right] \\
  -&\frac{B_p^2R}{nB^2}\Phi' \int d^3 v g
  -\frac{B_p^2R}{2nB\Omega} \frac{\p}{\p\psi} \int d^3 v v_\perp^2
  g 
\end{aligned}
\label{eq:Vt}\raisetag{3\baselineskip} 
\end{align}
where a prime denotes a $\psi$ derivative and $p=nT$.
The above expressions are accurate to
first order in $\delta B_p/B $, and thus include corrections due to
the gyrophase dependent part of the distribution function
$\tilde{f} \approx -\vec{\rho}\cdot \nabla f$,  where $\vec{\rho}
= \Omega^{-1}\vec{b} \times \vec{v}_\perp$ is the gyroradius vector. 
In the global theory, $\tilde{f}$ contributes with a $-\vec{\rho}\cdot \nabla g$ term, which gives the additional corrections to the diamagnetic and $E\times B$
flows on the second rows of the above equations.
Equation \eqref{eq:Vp} was calculated
in \nouncite{landreman12} (note that $g$ in \nouncite{landreman12} is
defined differently), and \eqref{eq:Vt} is calculated
analogously. We note, that while all terms are comparable
in \eqref{eq:Vp}, the global corrections to $V_t$ are small in $B_p/B$
compared to the usual expression (the first line of \eqref{eq:Vt});
nevertheless, we keep them for completeness.
  
Since all the terms in $V_p$ are proportional to $B_p/B$, we factor
out this trivial poloidal dependence and define the
\emph{poloidal flow coefficient}
\begin{equation}
k_p = \frac{Ze\langle B^2 \rangle}{IB_p} \left(\frac{\d T_0}{\d \psi}\right)^{-1} V_p,\label{eq:kp}
\end{equation}
which reduces to the conventional flux
function \emph{parallel flow coefficient} in the local
limit \cite{landreman12} (assuming the lowest order
distribution to be a flux function).

\subsubsection{\label{sec:flowCPhi}Fixed $\Phi$ profile flows}

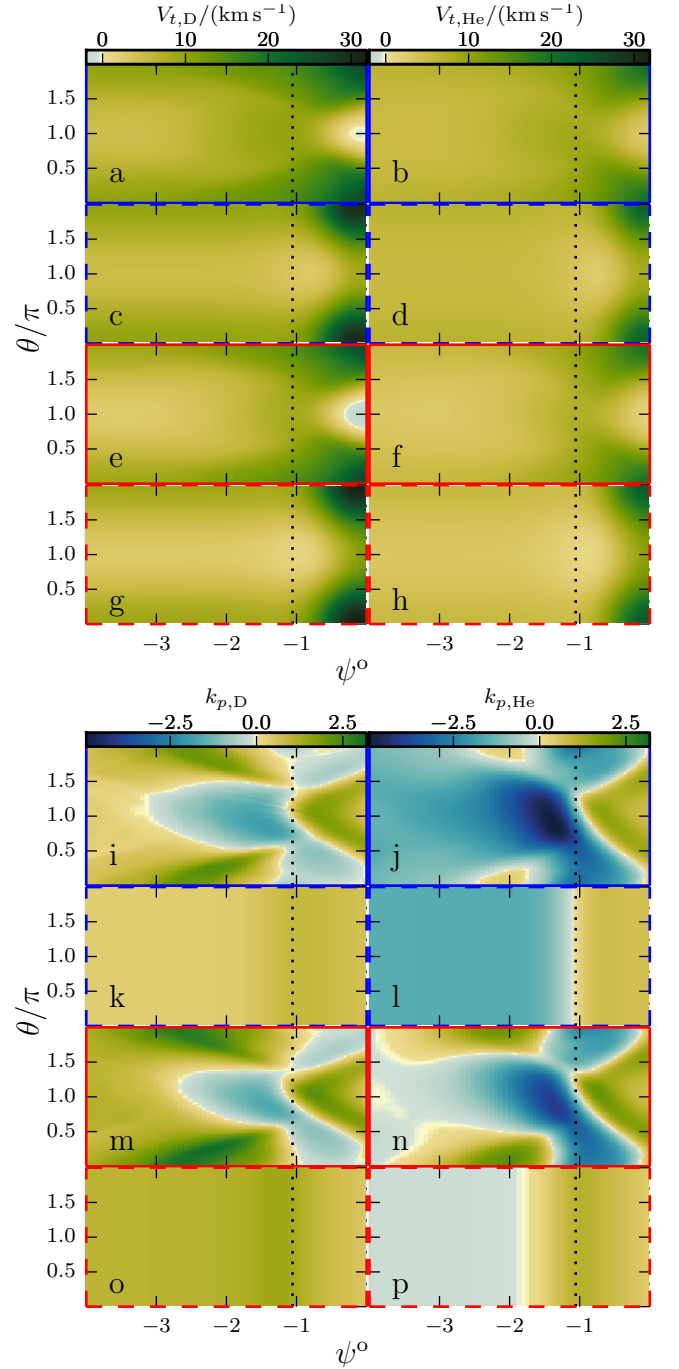
\begin{figure}
\begin{subfigure}[b]{1.0\columnwidth}
  \includegraphics{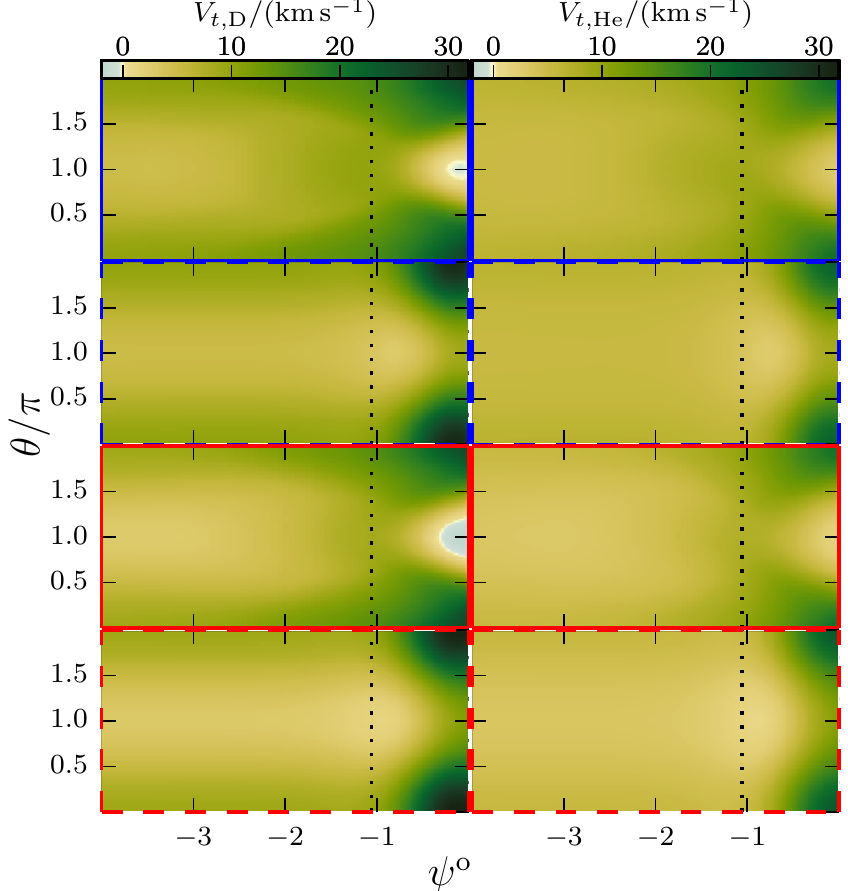}
\put(-207,190){\large a}
\put(-100,190){\large b}
\put(-207,136){\large c}
\put(-100,136){\large d}
\put(-207,82){\large e}
\put(-100,82){\large f}
\put(-207,28){\large g}
\put(-100,28){\large h}        
  \end{subfigure}
\begin{subfigure}[b]{1.0\columnwidth}
  \includegraphics{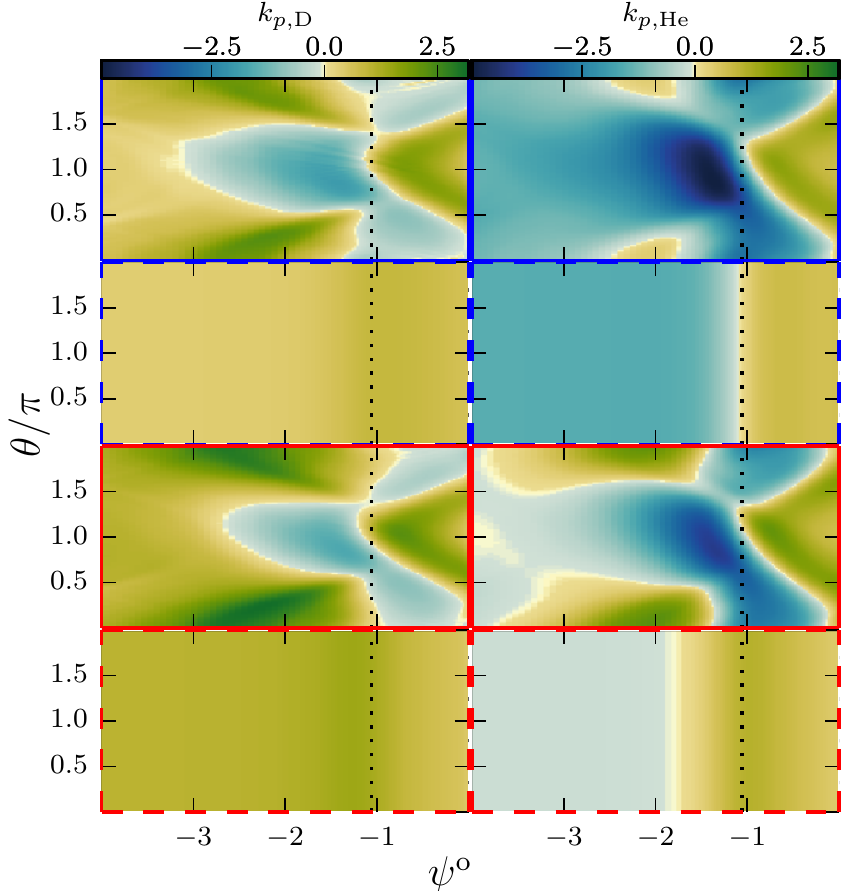}
\put(-207,190){\large i}
\put(-100,190){\large j}
\put(-207,136){\large k}
\put(-100,136){\large l}
\put(-207,82){\large m}
\put(-100,82){\large n}
\put(-207,28){\large o}
\put(-100,28){\large p}
\end{subfigure}
\caption{\label{fig:flow_d}D and He toroidal flows (a-h) 
and poloidal flow coefficients (i-p) for bulk deuterium (a-d, i-l,
  \Dbulk{}) and bulk helium (e-h, m-p, \Hebulk{}) plasmas, in the fixed $\Phi$
  scan. Dashed frames (c-d, g-h, k-l, o-p) are local
  results.}
\end{figure}

Using \eqref{eq:Vp}-\eqref{eq:kp} and the fixed $\Phi$ input profiles in \autoref{fig:dHeinputs}, we obtain
the ion flows, $k_p$ and $\hat{V}_t$, displayed in \autoref{fig:flow_d}.

For the toroidal flows -- even though the terms in the second 
line of \eqref{eq:Vt} are negligible -- global effects have an impact
through the modifications to $g$. Comparing the global toroidal flow
results, \autoref{fig:flow_d}a,b,e and f, to the corresponding local
ones, \autoref{fig:flow_d}c,d,g and h, the most striking difference is
that close to the LCFS the toroidal flow changes sign at the high-field side (HFS, $\theta=\pi$).
As a general trend seen at all poloidal locations the global toroidal
flows are elevated in the core plasma, and reduced in the pedestal.

The poloidal flow coefficients are flux functions in the local case, as
seen in \autoref{fig:flow_d}k,l,o and p. However, the corresponding
global results exhibit complex radial-poloidal features, shown
in \autoref{fig:flow_d}i,j,m and n. In particular, significant
poloidal variations appear in the flows, that include sign changes. 

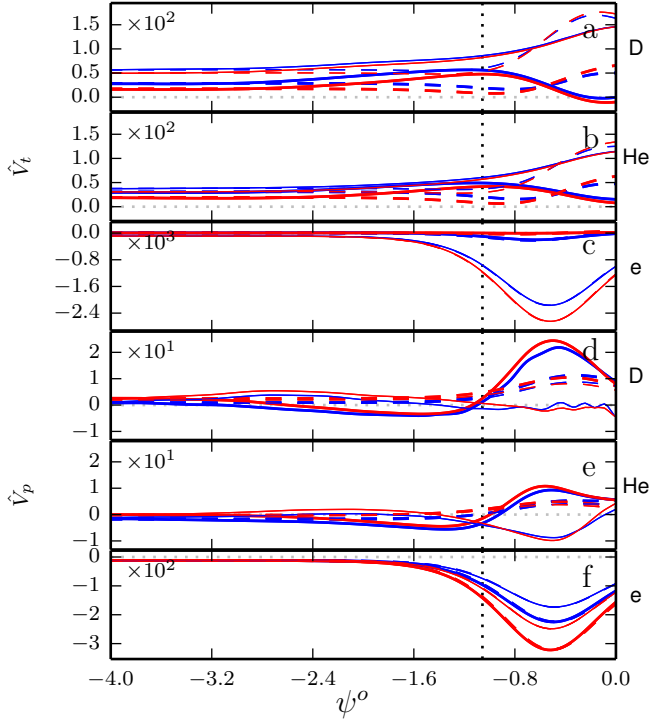
\begin{figure}
  \includegraphics{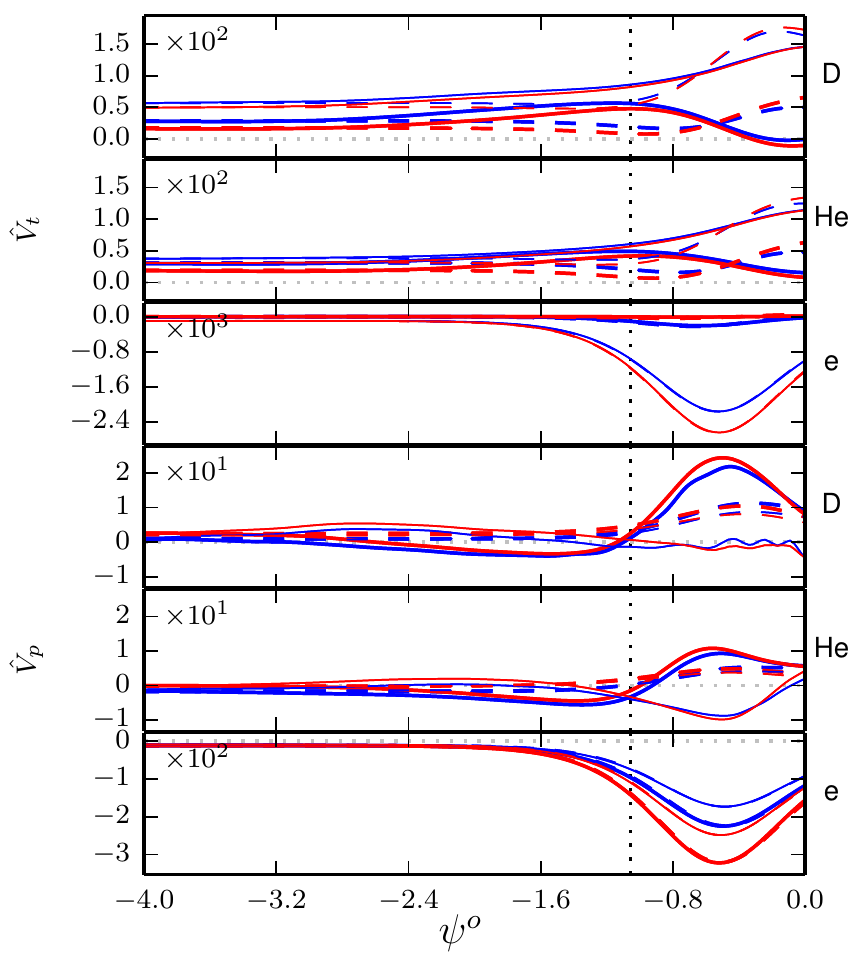} 
\put(-25,258){\large a} 
\put(-25,216){\large b} 
\put(-25,176){\large c} 
\put(-25,136){\large d} 
\put(-25,92){\large e} 
\put(-25,50){\large f} 
\caption{\label{fig:VpInOut}Toroidal (a-c)
  and poloidal (d-f) flows at the LFS ($\theta=0$, thin
  lines) and HFS (thick, $\theta=\pi$), in the fixed
  $\Phi$ scan. Main species is deuterium (\Dbulk{} lines) or helium
  (\Hebulk{}). Solid (dashed) lines correspond to global (local) results.}
\end{figure}

The high field side (HFS, $\theta=\pi$, thick lines) and low field side (LFS, $\theta=0$, thin lines)
toroidal and poloidal flows are shown in \autoref{fig:VpInOut}a-c and
d-f, respectively. The electron flows are mostly local due to their
small orbit width and our low flow ordering. However, for ions,
changing from local (dashed lines) to global (solid lines) simulations
leads to important changes. For instance, in the global case, while
the LFS toroidal flows monotonically increase for both ion
species throughout radial range plotted, their HFS
counterparts decrease in the pedestal, and even change sign for
deuterium.  On the other hand, for a given ion species, the effects of
changing its role from bulk to impurity are rather small, and they are
similar in the global and local simulations (\Dbulk{} curves: D bulk;
\Hebulk{}: He bulk). In the weakly collisional limit of the local theory,
even those flow contributions, which are ultimately caused by
collisions, become independent of collision frequency. The observed
weak variation with changing ion concentration seen here is the result
of these cases not being asymptotically collisionless and due to
inter-species coupling (as confirmed by simulations with artificially
increased collisionality, not shown here). This feature remains valid
in the global case as well.

\subsubsection{\label{sssec:cne}Fixed $n_e$ profile flows}
In contrast to the fixed $\Phi$ scenario, in the fixed $n_e$ scenario
both $\Phi$ and the ion density profiles change with ion composition,
which is expected to be reflected in the ion flows. Since these
profile changes are limited to the pedestal region, this is where
corresponding effects are expected in the local theory. All
modifications to local flows outside the pedestal are the result of
changes in collisionality, and thus should be similar in the two
similarity classes.  In the global theory the effect of the changes in
the pedestal profiles will propagate outside the pedestal due to the
radial coupling of $g$. It is important that, although in the pedestal
$\Phi$ and the ion density gradients change in the scan, the $\eta$
gradient is not changed due to the construction of our model
profiles. This means that any differences observed are not due to
changes in $\partial_\psi f_M$, but due to changes in the finite orbit
width effects (the magnitude of the radial electric field is reduced
with increasing He concentration).

\begin{figure}
  \includegraphics{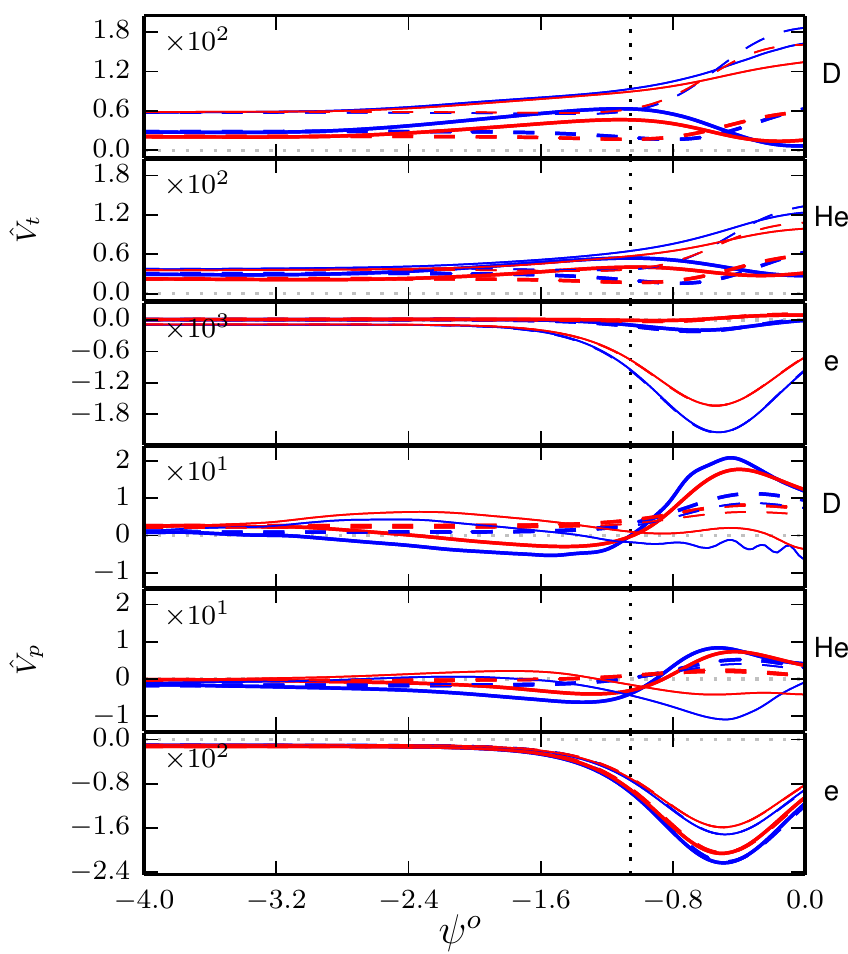} 
\put(-25,258){\large a} 
\put(-25,216){\large b} 
\put(-25,176){\large c} 
\put(-25,136){\large d} 
\put(-25,92){\large e} 
\put(-25,50){\large f} 
\caption{\label{fig:neVpInOut}Toroidal (a-c)
  and poloidal (d-f) flows at the LFS ($\theta=0$, thin
  lines) and HFS (thick, $\theta=\pi$), in the fixed
  $n_e$ scan. Main species is deuterium (\Dbulk{} lines) or helium
  (\Hebulk{}). Solid (dashed) lines correspond to global (local) results. }
\end{figure}

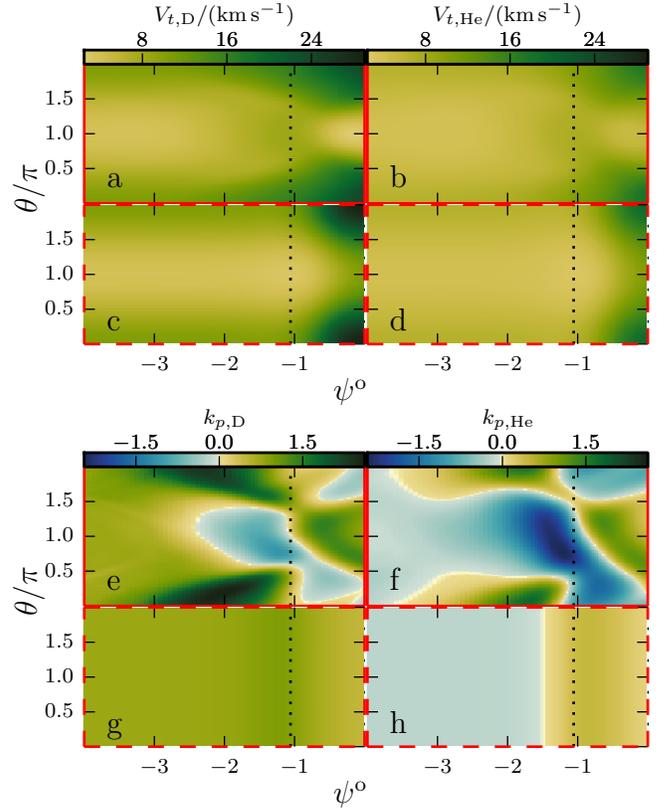
\begin{figure}
\begin{subfigure}[b]{1.0\columnwidth}
  \includegraphics{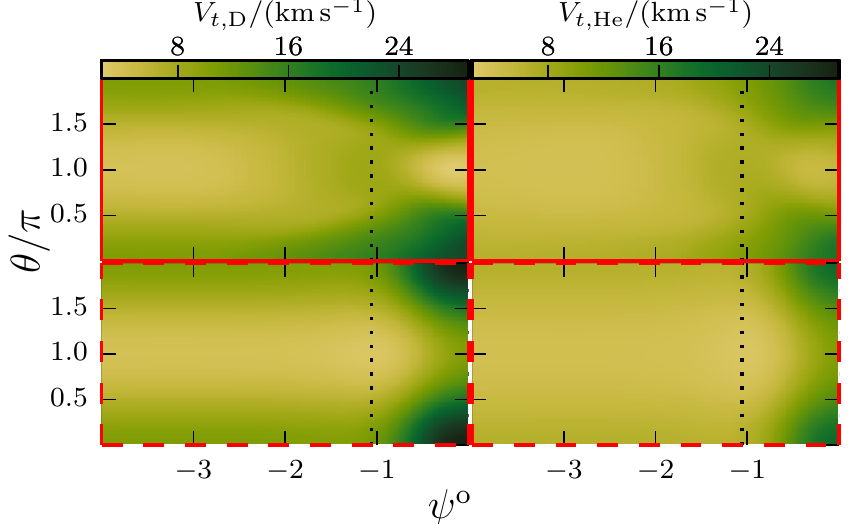}
\put(-207,82){\large a}
\put(-100,82){\large b}
\put(-207,28){\large c}
\put(-100,28){\large d}        
  \end{subfigure}
\begin{subfigure}[b]{1.0\columnwidth}
  \includegraphics{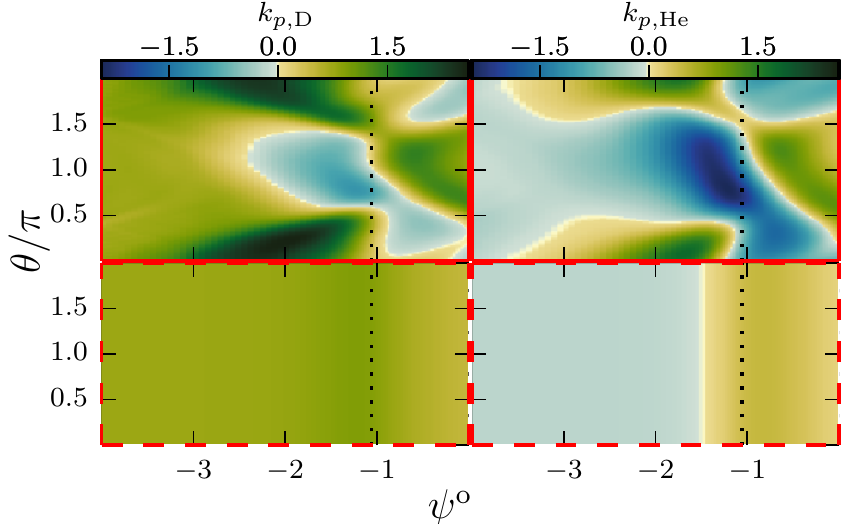}
\put(-207,82){\large e}
\put(-100,82){\large f}
\put(-207,28){\large g}
\put(-100,28){\large h}
\end{subfigure} 
\caption{\label{fig:neflow_d}Toroidal
      flows(a-d) and poloidal flow coefficients (e-h) in a bulk helium
      plasma similar to \autoref{fig:flow_d}(e-h) and (m-p), but
      for fixed $n_e$ profiles. } 
\end{figure}

The flows depicted in \autoref{fig:neVpInOut} and
\autoref{fig:neflow_d} do not reveal a striking variation with plasma
composition.  If we compare the LFS ion flows in the pedestal (thin
lines, sub-figures: a-b, d-e; at $\psio=0$) in \autoref{fig:neVpInOut}
to the corresponding fixed $\Phi$ figure, \autoref{fig:VpInOut}, we
indeed see a stronger effect of exchanging bulk and impurity species,
although the two different scans produce curves with the same
qualitative features. On the other hand, the HFS flows are not
affected as strongly. Overall, the effect of changing the helium
concentration is small in both scans. Thus, for our low
collisionality, slight variations in density profiles do not matter
much. The similarities between the two scans can also be verified by
looking at the full poloidal dependence of the flows, shown in
\autoref{fig:neflow_d} for the bulk helium simulation.

Since we have established that the electron dynamics is not too much
affected by our low-flow ordered ions, and the $n_e$ profile is held
fixed, we expect the poloidal electron dynamics to exhibit only a
modest change in the scan. Comparing the differences between the bulk
D (\Dbulk{}) and bulk He (\Hebulk{}) curves in
between \autoref{fig:neVpInOut}f and \autoref{fig:VpInOut}f, we find
that this is indeed the case. On the other hand, the toroidal flow of
electrons changes in the fixed $n_e$ scan, as the radial electric
field varies with ion composition.

\subsubsection{\label{sssec:dff}Interplay of the poloidal and radial dynamics}

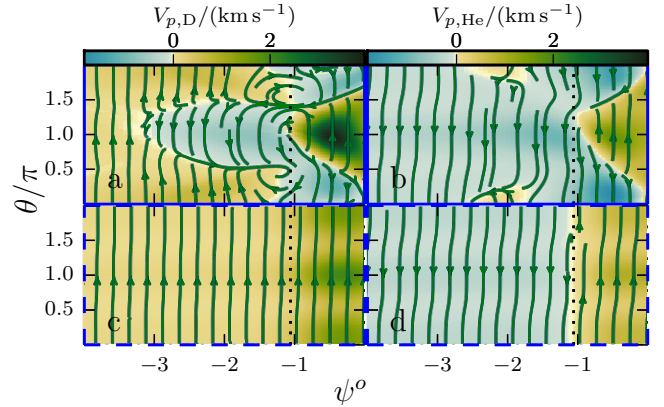
\begin{figure}
  \includegraphics{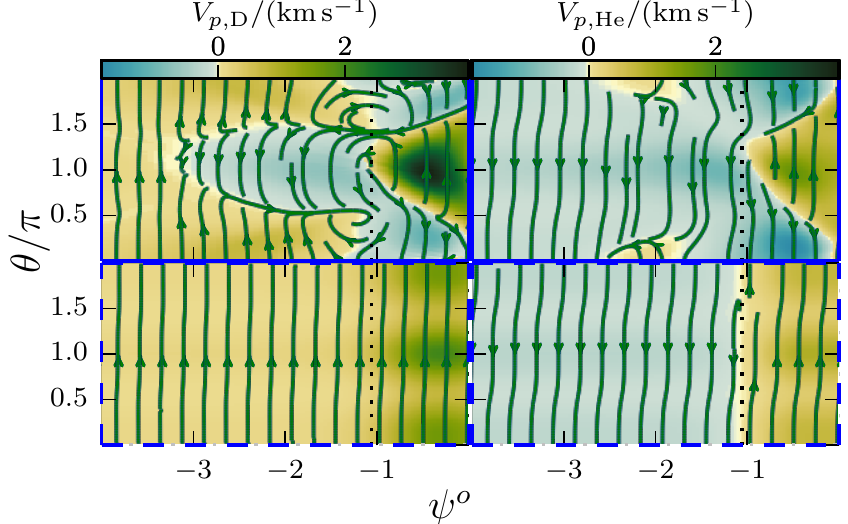}
\put(-207,82){\large a}
\put(-100,82){\large b}
\put(-207,28){\large c}
\put(-100,28){\large d}
\caption{\label{fig:stream} A stream-plot of fluxes in the radial-poloidal 
plane, overlaid on a color map of the poloidal flows, in a bulk
deuterium plasma. Deuterium (a,c) and helium (b,d) species in global
(a-b) and local (c-d) simulations.}
\end{figure}

In the global simulations we found that the poloidal flow can change sign
between different poloidal locations. This is possible because the
divergence of the radial flux can make the total radial-poloidal
fluxes divergence free. (We use the term ``flux'' instead of ``flow'',
because the radial variation of the density is not negligible in our
simulations.) It is therefore instructive to consider the full
radial-poloidal structure of the fluxes. \autoref{fig:stream} shows a
stream plot of the fluxes in the poloidal and radial directions
overlaid on top of the poloidal flows, for the same case
as shown in \autoref{fig:flow_d}i-l.
To account for the vastly different length-scales in the
$\nabla\theta$ and $\nabla\psi$ direction, the fluxes are normalized to typical
pedestal and poloidal lengths, specifically
$(\vec{\Gamma}\cdot \vec{\psi}/(\Delta
r),\vec{\Gamma}\cdot \vec{\theta}/(2\pi a))$, where $\Delta r$
is the pedestal width in meters, $a$ the minor radius on the outboard side and $\vec{\theta}$ and $\vec{\psi}$ are unit vectors in the $\nabla\theta$ and $\nabla\psi$ direction, respectively.

As seen in the lower panels of \autoref{fig:stream}, the dynamics of
the local simulations is rather simple: the small radial fluxes are
superimposed on weakly varying poloidal fluxes. In contrast, in the
global simulations we find a much richer pattern of fluxes, with
stagnation points and vortices in the radial-poloidal
plane. 
Sufficiently far from the pedestal region the flows approach
their local behavior. The more complex radial flow patterns are not
completely localized to the pedestal, but extend a few orbit widths
into the core region, which is more visible for the deuterium species.

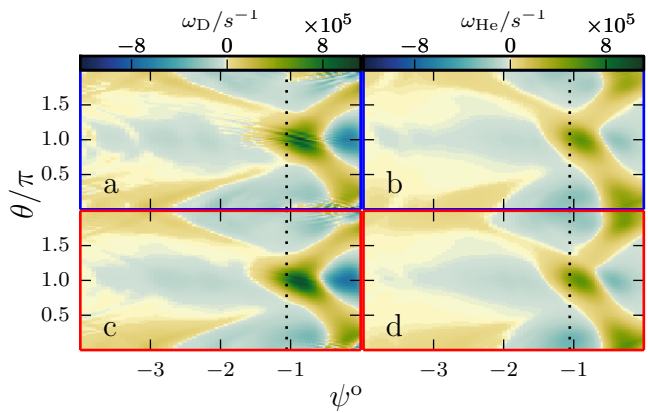
\begin{figure}
  \includegraphics{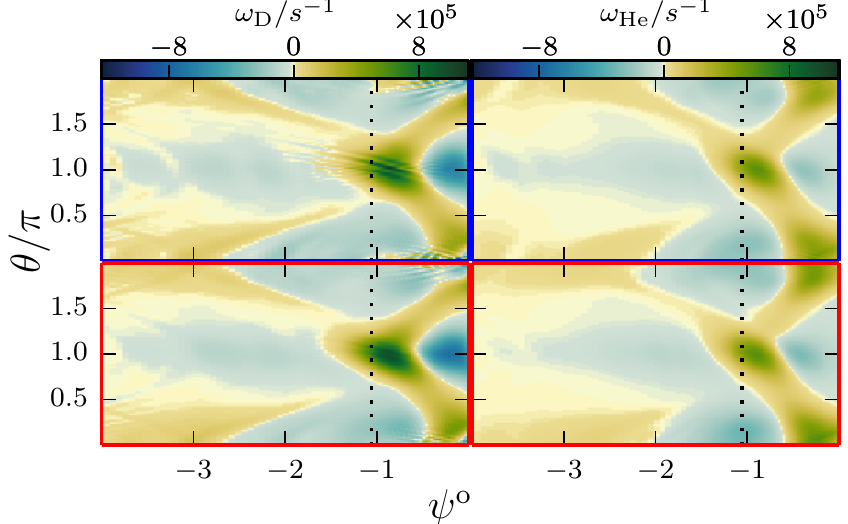}
\put(-207,82){\large a}
\put(-100,82){\large b}
\put(-207,28){\large c}
\put(-100,28){\large d}  
  \caption{\label{fig:vorticity} Deuterium and helium vorticity (defined in \eqref{eq:vorticity}) for bulk deuterium (a-b,
  \Dbulk{}) and bulk helium (c-d, \Hebulk{}) plasmas, from the global simulations in the fixed $\Phi$ scan.}
\end{figure}

The vortices in \autoref{fig:stream} depend on the values of radial and poloidal fluxes, and are thus sensitive to the slight shifts in flows observed by changing plasma composition. To only consider the spatial variations of the fluxes, we evaluate the vorticity, defined as
\begin{equation}
\omega = \frac{1}{\bar{n}}\hat{\varphi} \cdot \nabla \times \vec{\Gamma} \sim \frac{1}{\bar{n}} \left(\frac{\d \Gamma_p}{\d r} - \frac{1}{a}\frac{\d \Gamma_r}{\d \theta}\right). \label{eq:vorticity}
\end{equation}
The vorticities corresponding to \autoref{fig:flow_d} are displayed in
\autoref{fig:vorticity}. Just like the flows, the vorticities are
largely unaffected by exchanging bulk and impurity species, and both
helium and deuterium display similar structures.  In particular, both
species display ``V''-like arms of high vorticity in the pedestal. The
apparent differences between the He and D flows in
\autoref{fig:flow_d} may be due to constant flows that mask the radial
and poloidal variations, which are more sensitive to global physics.
Similarly, sloped vorticity structures extend from the pedestal into
the core, and reveal global effects reaching far into the core.  It is
interesting to note that the slope of the arms is different between
the different species, so that the helium vorticity structure extends
over all poloidal angles within the width of the pedestal. This
behavior is consistent with helium flows approaching their local
values sooner due to their smaller orbit width.

\begin{figure}
  \includegraphics{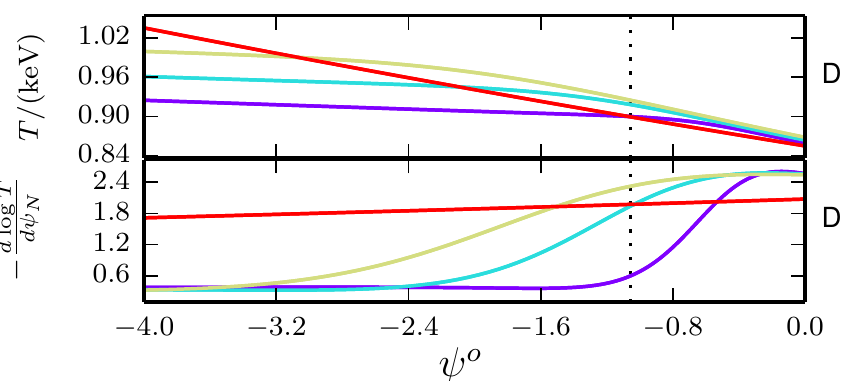} 
\put(-195,68){\large a}
\put(-195,25){\large b}
\caption{\label{fig:T_Tt}
      Temperature (a) and logarithmic temperature gradient (b)
      profiles in the transition-length scan. Violet: baseline
      temperature profile in \autoref{fig:dHeinputs}; cyan, a
      transition length doubled; yellow, transition length tripled;
      red, no transition in temperature gradients.}  
\end{figure}

The divergence of the radial fluxes is strongly affected by a radial
variation in the diamagnetic flux, and thus it is expected that
radially global flow effects are localized to regions where the
density or the driving gradients abruptly change. For our model
profiles, the density drops inside the pedestal, and the ion
temperature gradient rapidly increases at the pedestal top (as
expected in a real pedestal). In order to demonstrate the effect of
the location of changes in diamagnetic flow strength, we modify our
ion temperature profiles. In a scan, we increase the radial length
scale over which the logarithmic temperature gradient transitions from
its core value to its pedestal value; see the corresponding
temperature profiles and logarithmic gradients
in \autoref{fig:T_Tt} (transition length scale increases from violet
to yellow). 

\begin{figure}
  \includegraphics{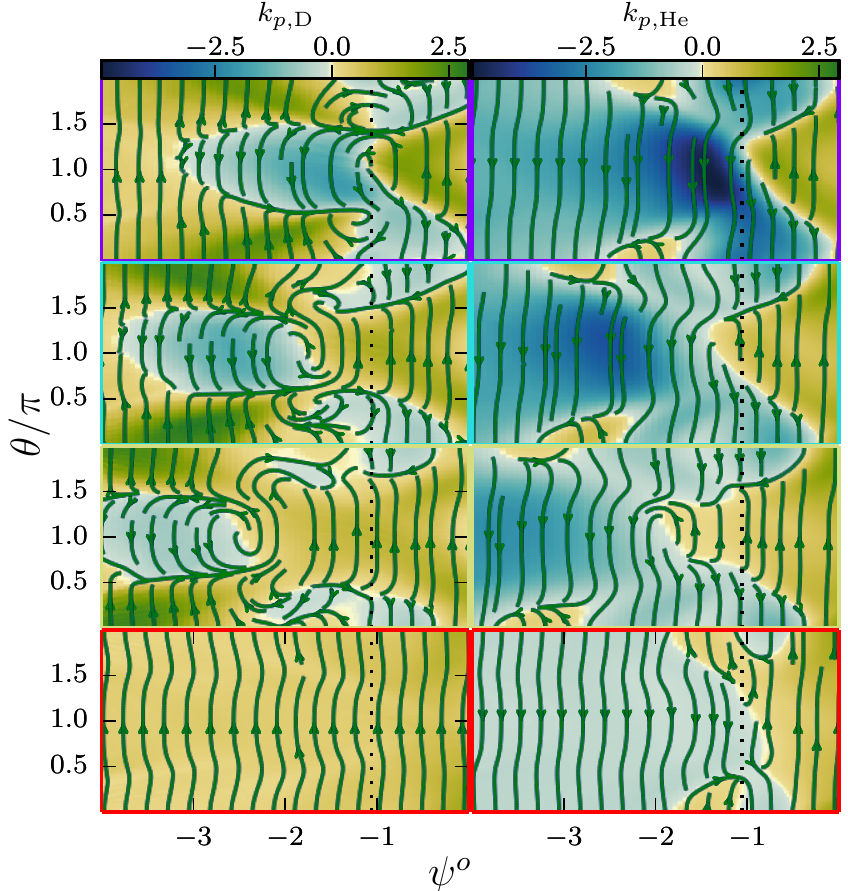}
\put(-207,190){\large a}
\put(-100,190){\large b}
\put(-207,136){\large c}
\put(-100,136){\large d}
\put(-207,82){\large e}
\put(-100,82){\large f}
\put(-207,28){\large g}
\put(-100,28){\large h}
\caption{\label{fig:flow_Tt} Foreground: stream-plot of the
deuterium and helium fluxes in the radial-poloidal plane. Background:
$k_p$.  (a-b): Baseline temperature profile
in \autoref{fig:dHeinputs}; (c-d): transition length doubled; (e-f):
transition length tripled; (g-h): no transition. D (a,c,e,g ) and He
(b,d,f,h) ion species in a deuterium bulk plasma.}  
\end{figure}

Increasing the transition region has a twofold effect: the transition
becomes less abrupt, and the effective location where the transition
happens moves further outside the pedestal.  As a result, the radially
global effects in the flow structure become weaker and start to extend
further away from the pedestal, as illustrated
in \autoref{fig:flow_Tt}. We also consider a temperature profile with
no transition region (red curve) in \autoref{fig:T_Tt},
and \autoref{fig:flow_Tt}g and h. This shows much weaker, but still
visible, deviations from the local simulation in terms of flow
patterns. However we have to interpret the ``no-transition'' results
carefully, since in this case the sources were non-negligible even
close to the radial boundaries. Nevertheless, it also underlines the
importance of the changes in the ion temperature gradient in driving
unexpected poloidal flow patterns.

\begin{figure}
\includegraphics{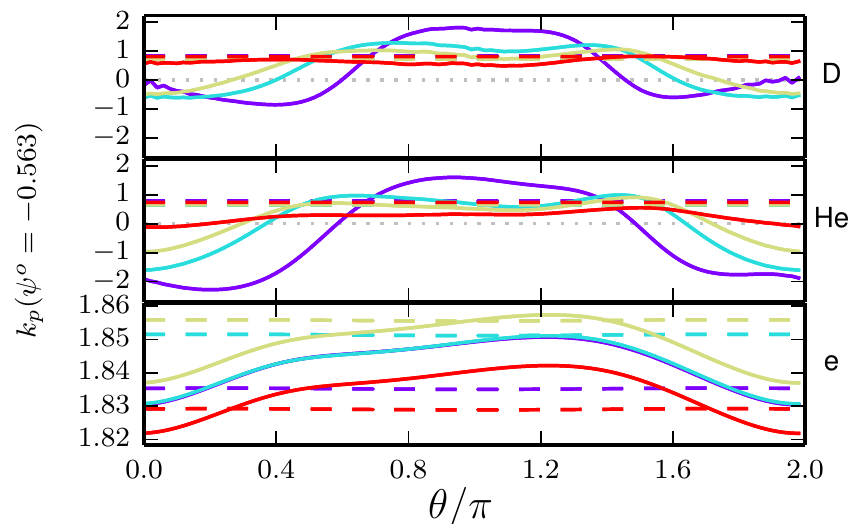} 
\caption{\label{fig:k_p_Tt}The $k_p$ corresponding to 
\autoref{fig:flow_Tt}a,c,e and g, in the middle of the pedestal ($\psio=-0.563$) for local (dashed lines) 
and global (solid)  simulations.}
\end{figure}

Finally, we consider the poloidal structure of the poloidal flows in
the middle of our pedestal region for the various temperature profiles
of \autoref{fig:T_Tt}. The poloidal flow coefficients at $\psio =
-0.563$ are shown in \autoref{fig:k_p_Tt} with the same color
coding. For the baseline case (violet line) there are substantial
poloidal variations of $k_p$ for both ion species, being higher (or
more positive) on the HFS, and lower (more negative) on the
LFS. The variation is not sinusoidal, and the local maxima
and minima (several of them) appear at poloidal locations away from
the mid-plane. With increasing transition length (cyan and yellow
curves), the poloidal variation becomes milder, while the minima move
to the outboard mid-plane and merge, and the maxima move towards the
upper and lower parts of the flux surface. In the no-transition case
(red curve), the poloidal variation is weak, but still present, and the
global value of $k_p$ is lower than the local one. For electrons
(\autoref{fig:k_p_Tt}c), the poloidal flows essentially remain local
(note the magnified y-axis scale).

\subsection{\label{sec:flux}Fixed $\Phi$ radial flows and fluxes}
\begin{figure}
  \includegraphics{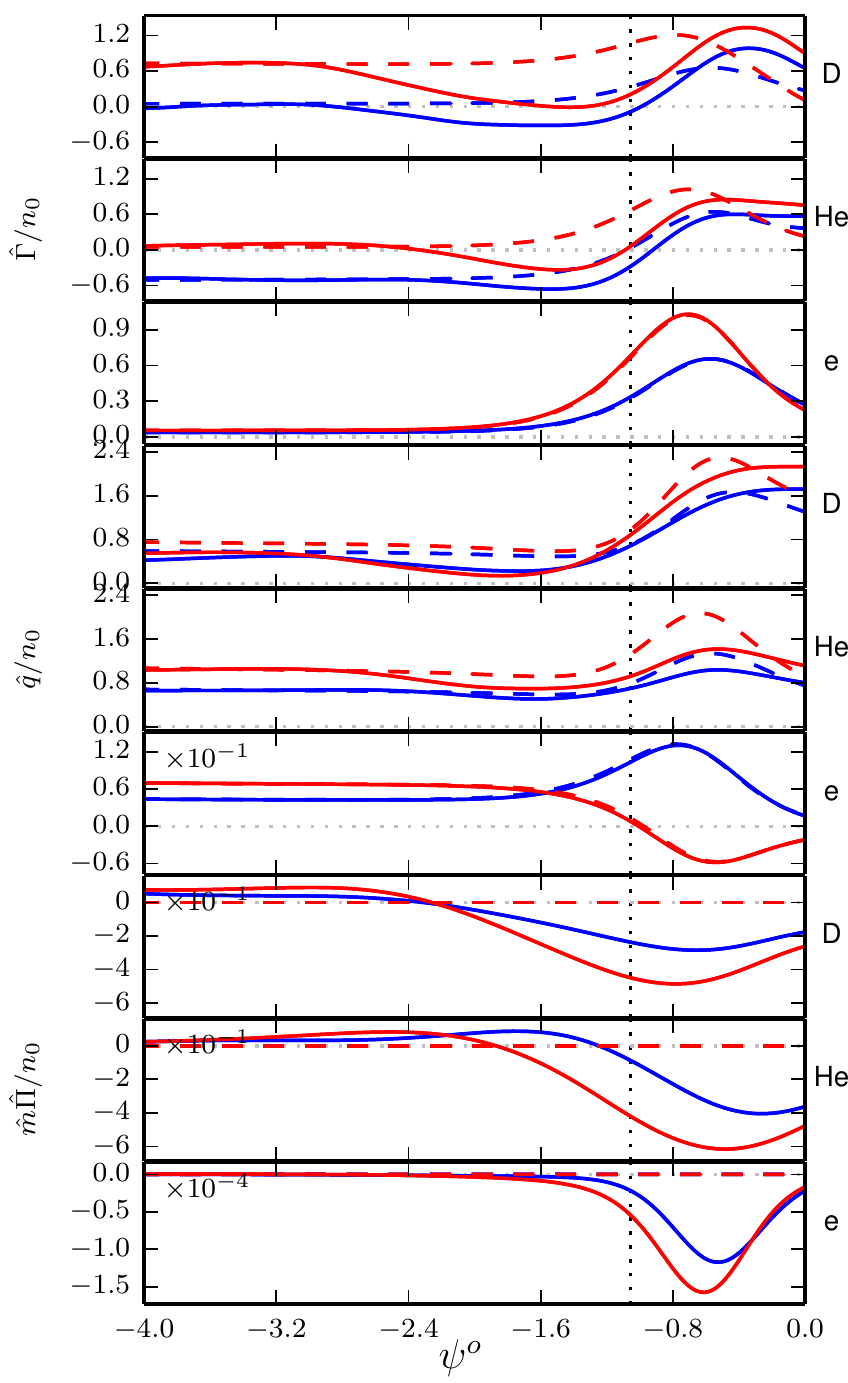}
\put(-25,384){\large a}
\put(-25,342){\large b}
\put(-25,300){\large c}
\put(-25,256){\large d}
\put(-25,216){\large e}
\put(-25,174){\large f}
\put(-25,118){\large g}
\put(-25,74){\large h}
\put(-25,40){\large i}
    \caption{\label{fig:dHeflux} Particle (a-c), heat (d-f) and
      momentum (g-i) fluxes for bulk D (\Dbulk{} curves) and He (\Hebulk{}) plasmas in the fixed $\Phi$ scan.
      Solid (dashed) lines are global (local) results.}
  \end{figure}

\begin{figure}
  \includegraphics{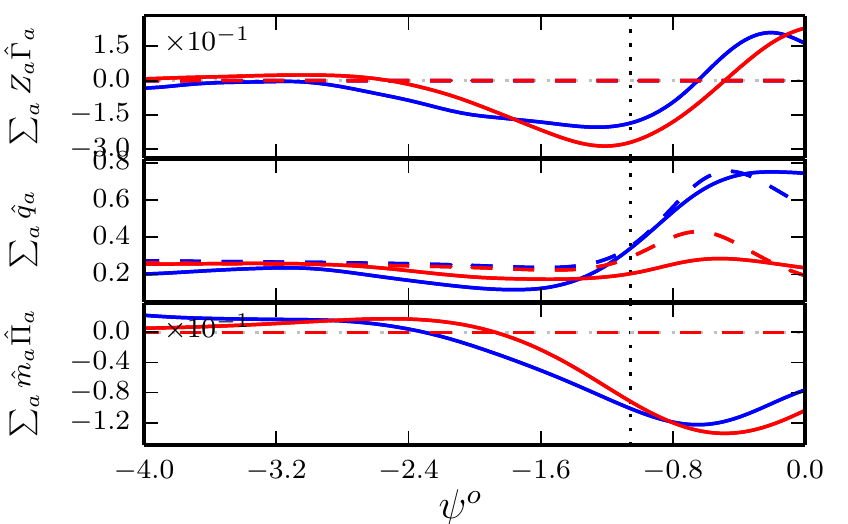}
\put(-25,136){\large a}
\put(-25,88){\large b}
\put(-25,52){\large c}
  \caption{\label{fig:dHeTotFlux} Current (a) and total heat (b) and momentum
    (c) flux, for bulk D (\Dbulk{} curves) and He (\Hebulk{}) plasmas in the fixed $\Phi$ scan.
    Solid (dashed) lines are global (local) results.}
\end{figure}

To get a clearer picture of the effects of helium concentration on
transport, we calculate the total radial particle, conductive heat and
toroidal momentum fluxes in the fixed $\Phi$ scan. These are displayed in
\autoref{fig:dHeflux}, normalized by the core species density
$\hat{n}_{0}$ to assist the comparison of the trace and non-trace
scenarios.

From \autoref{fig:dHeflux}, we can make a few general observations.
Firstly, the electron fluxes are practically local by virtue of their
small orbit width and large flows, so the electron global
and local curves almost overlap (except for the momentum flux
  \autoref{fig:dHeflux}i, which has no physical relevance as it is
  small in the electron-to-ion mass ratio -- we show it nevertheless for
  completeness). Since electrons are often omitted from neoclassical
  simulations, it deserves mention that they can develop a substantial
  particle flux inside the pedestal. This is the result of the large
  electron temperature gradient, and it is also present in local
  simulations. 

Secondly, as seen in \autoref{fig:dHeflux}a-b and
\autoref{fig:dHeflux}d-e, the global ion fluxes in the near-pedestal
core tend to be reduced compared to the corresponding local
fluxes. Modifications to the ion heat flux compared to the
conventional local theory have been predicted by analytical
models\cite{kagan08,pusztai10,catto11,catto13} retaining the
$\vec{v}_E\cdot \nabla g$ term, however in our simulations the radial
coupling from the $\vec{v}_m\cdot \nabla g$ also plays an important
role in setting the radial fluxes. Here, the modifications are
especially notable for the particle fluxes \autoref{fig:dHeflux}a-b,
which change sign compared to the local results in the near-pedestal
core. The width of the affected region scales with the orbit width and
is thus larger for D then He; accordingly, the fluxes reach local
values further away from the pedestal in the D plasma. Such
``overshoot'' behavior is also seen in the momentum fluxes
\autoref{fig:dHeflux}g-h, which instead are increased in the
near-pedestal core compared to their local value (that is zero).

The flux surface average species-summed toroidal angular momentum
balance states that the named quantity changes in time due to a
divergence of the radial momentum transport ($\sim\d_\psi\sum_a
\hat{m}_a \hat{\Pi}_a$), a torque corresponding to any radial currents
($\sim \sum_a Z_a \hat{\Gamma}_a$), or momentum sources. In our
\emph{steady state} simulations these three contributions should add
up to zero. However, in the calculations presented here there are no
momentum sources, thus any momentum transport requires the existence
of a radial current. This is indeed the case:
\autoref{fig:dHeTotFlux}a and c show the corresponding finite radial
currents and toroidal angular momentum transport,
respectively. (Recall that we do not enforce the ambipolarity of
fluxes, and the radial electric field is not solved for in our
simulations -- but is an input -- while the flows are outputs.) This
explains why the particle fluxes are below the local values on one
side of the pedestal, and above on the other side: the current must
integrate to zero over the entire domain for the momentum transport to
approach its local value (i.e. vanish) far from the pedestal. We note
that although there are non-intrinsically ambipolar processes in the
pedestal which could balance our radial currents (due to orbit losses,
ripple fields, etc.), the radial current we observe is not a necessary
feature, but rather a consequence of not allowing for momentum
sources. In Appendix~\ref{sec:momentum} we demonstrate that radial
currents can be replaced by momentum sources, and it does not have a
significant effect on the flow structures.

\begin{figure}
  \includegraphics{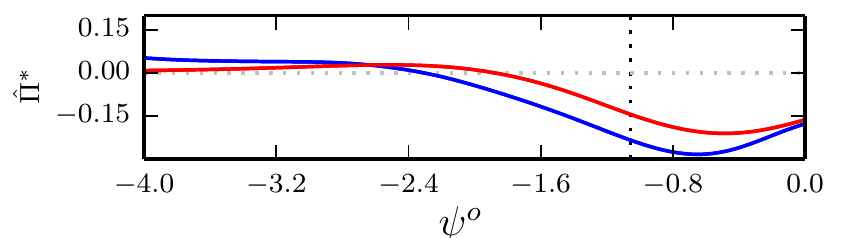}
  \caption{\label{fig:prandtl} Total momentum flux in \autoref{fig:dHeTotFlux}c normalized to be a proxy for the effective Prandtl number in the pedestal \eqref{eq:prandtl}, for the fixed $\Phi$ scenario.}
\end{figure}

From \autoref{fig:dHeTotFlux}c, we see that the total momentum flux
has roughly the same peak value in both D and He plasmas, although the
shape of the curve is wider in the bulk deuterium simulation, again
due to the larger orbit width. We emphasize that the finite
  momentum transport observed here is purely a radially global
  effect. In local theory the momentum transport is a small, higher
  order quantity. When $\delta_X$ is not small for all input profiles
 $X$, momentum transport is not small in general. To demonstrate that
  the magnitude of the momentum transport we observe can be
  experimentally relevant, we normalize the momentum flux so that it becomes
  a proxy for the effective Prandtl number in the pedestal\cite{tala2007ppcf}
  \begin{equation}
    \hat{\Pi}^{*} = \sum_a \hat{m}_a \hat{\Pi}_a \left\{\frac{ \hat{n}_\text{i} d_\psi \hat{T}_\text{i}}{ \hat{q}^{}_\text{i}\Delta} \frac{1}{ \hat{m}_\text{i}  d_\psi \left(\langle \hat{V}_{\text{t,i}} \rangle \hat{n}_\text{i}\right)}\right\}_\text{ped} \label{eq:prandtl}.
  \end{equation}
  The expression with the ``ped'' subscript is evaluated in the middle
  of the pedestal ($\psio = -0.563$), and ``i'' denotes the bulk ion
  species\footnote{Note that \nouncite{tala2007ppcf} defines the
    momentum flux in terms of $mv_t$, with $v_t$ being the toroidal
    velocity, while we use toroidal angular momentum over mass $R
    v_t$.}.  This $\Pi^*$ is displayed in \autoref{fig:prandtl}, for
  the fixed $\Phi$ simulations.  We find that the effective Prandtl
  number reaches about $0.35$ and $0.24$ for D and He bulk plasmas,
  respectively. This is comparable to experimentally observed
  effective Prandtl numbers in the plasma core of
  JET\cite{JETmomentumPinch} and
  KSTAR\cite{kstarToroidalRotation}. Since experimental ion heat
  transport can be comparable to the neoclassical predictions in a
  pedestal\cite{viezzerIAEANF2017}, assuming that our results
  extrapolate to large temperature gradient, we may expect finite
  orbit width effects to significantly contribute to the total
  momentum transport in a pedestal\footnote{Charge-exchanging neutrals
    coupled to the neoclassical ion distruburion is another mechanism
    where collisional physics is relevant to the momentum transport in
    the edge\cite{FulopNeutralPRL,omotani16}.}.

\begin{figure}
  \includegraphics{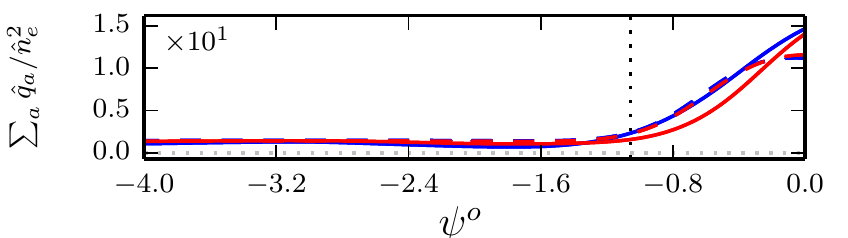}
    \caption{\label{fig:dHeqflux_over_ne} The total heat flux in
      \autoref{fig:dHeTotFlux}b divided by $\hat{n}_e^2$.}
  \end{figure}

Next we consider bulk species effects on the total neoclassical
conductive heat flux, shown in \autoref{fig:dHeTotFlux}b.  Its peak
value is reduced by a factor $4$ when going from a deuterium to a
helium plasma. This is consistent with the $q \propto n_e^2$ scaling
from the local banana-regime analysis\cite{helander2005}, and
compensating for this scaling yields \autoref{fig:dHeqflux_over_ne},
where the different simulations have similar peak values.

We also observe that the helium heat flux is lowered compared to local
predictions (\autoref{fig:dHeflux}e), whereas the deuterium heat flux
is comparable to the local values (\autoref{fig:dHeflux}d). This is due to
the fact that the orderings \eqref{delta} make the helium density
pedestal sharper than the deuterium density pedestal, which results in
a lower $n^{}_\text{He}$ further out in the pedestal, and thus a reduced
heat flux, from $\hat{q}\propto \hat{n}^2$. Since the ion heat fluxes
in the global simulations peak further out compared to the local
simulations (\autoref{fig:dHeflux}d-e), the global peak values of
$\hat{q}$ are also affected by the
reduced $n_\text{He}$ in the buffer region, which may be a
  contributing factor to the reduction of the global
$q_\text{He}$ compared to the local value.

\section{Conclusions}
\label{sec:disc}
We have studied finite orbit width effects on the neoclassical
toroidal and poloidal flows and cross field fluxes, in density
pedestals of deuterium-helium mixture plasmas.  In a radially global
treatment we allow for ion orbit width scale radial density variations
and strong radial electric fields, as long as the ions are
electrostatically confined and are characterized by subsonic flow
speeds.  The deuterium-helium ratio scans were performed keeping
either the electrostatic potential variation or the electron density
profile fixed, leading to surprisingly similar results.

The perturbed neoclassical distribution is modified compared to the
radially local treatment, since magnetic and $E\times B$ drift
contributions to the advection of the perturbed distribution need to
be retained. In addition, non-standard terms appear in the expressions
for the flows, corresponding to an $E\times B$ advection of poloidal
density perturbations and, more importantly, to the radial variation
of diamagnetic fluxes. The resulting poloidal flows exhibit complex
radial-poloidal features which vary on a small radial scale, including
poloidal sign changes of the poloidal flow. The main reason is that
the poloidally local radial particle fluxes are not divergence free in
isolation due to the sharp profile variations, which require poloidal
return flows to make the total fluxes divergence free. Such flow
structures are found to be sensitive to abrupt radial variations in
the ion temperature gradient, and can extend quite far into the core
if the ion temperature gradient transitions between its core and
pedestal value on a long radial scale.

The near-pedestal core values of the global neoclassical particle and
conductive ion heat fluxes are often reduced compared to the local
results, as a result of an overshoot of decreased fluxes away
from the pedestal. Inside the pedestal the heat fluxes are mostly
reduced compared to their local values. 

We observe a finite radial current, which at least partially arises as
no momentum sources were included in these simulations. However it can
also represent a physically meaningful charge separation process due
to finite orbit width effects, which in steady state needs to be
compensated by other non-intrinsically ambipolar processes. The
effects of replacing radial current with momentum sources is the
subject of ongoing investigation.

The sizable neoclassical toroidal angular momentum transport we
observe only appears in global theory. The momentum flux, when
normalized to represent an effective Prandtl number, takes on values
(few tens of percent) comparable to experimental values of the
effective Prandtl number observed in the plasma core. This is a
potentially important observation since the heat fluxes in the
pedestal can be dominated by the neoclassical ion heat flux. This
implies that if our results extrapolate to large ion temperature
gradients (where a full-f treatment is unavoidable), radially global
effects might account for a significant fraction of the momentum
transport in the inner region of the pedestal.

\acknowledgments The authors are indebted M. Landreman for support
with the PERFECT code and for enlighntening discussions. The authors
are also grateful to T. F\"{u}l\"{o}p and E. Highcock for discussions
and instructive comments on the manuscript.  IP and SB were supported
by the International Career Grant (Dnr.~330-2014-6313), JO and SN by
the Framework grant for Strategic Energy Research (Dnr.~2014-5392) of
Vetenskapsr{\aa}det.  The simulations used computational resources of
Hebbe at C3SE (project nr. C3SE2016-1-10 \& SNIC2016-1-161) and the
ARCHER UK National Supercomputing Service.

\appendix

\section{\label{sec:profiles} Model pedestal profiles and magnetic geometry}
Due to the constraints on the $T$ and $\eta$ profiles \eqref{delta},
we use simple model profiles for our simulations.

Specifically, an mtanh transition between
two constant gradient regions is implemented as
\begin{equation}
\begin{aligned}
X =& \frac{X_\text{ped}  + X_\text{SOL} + (a-b)w_\text{ped}/2}{2} \\
   &- \frac{X_\text{ped} - X_\text{SOL} + (a+b)w_\text{ped}/2}{2} \tanh{\left(r\right)} \\
   &+ \frac{a(\psi-\psi_0)\e^{-r} + b(\psi-\psi_0)\e^{r}}{\e^{-r} + \e^{r}},
 \end{aligned}
\end{equation}
where $X$ is a generic profile, $r=\frac{(\psi
- \psi_0)}{w_\text{ped}/2}$, $w_\text{ped}$ is the pedestal width, $a$
($b$) the core (SOL) asymptotic profile gradients; $\psi_0$ is the
position of the middle of the pedestal. Here ``SOL'' represents the numerical buffer region outside $\psio=0$.

The magnetic field is assumed not to vary notably over the pedestal
region, and we thus use a radially constant, local Miller geometry\cite{miller98}
with parameters: $\kappa=1.58$, $s_\kappa\equiv (r/\kappa)\d\kappa/\d
r=0.479$, $\delta = 0.24$, $s_\delta\equiv
(r/\sqrt{1-\delta^2})\d\delta/\d r=0.845$, $\p R/\p r =-0.14$,
$q=3.5$, $\epsilon\equiv r/R=0.263$, where $\kappa$ is the elongation,
$\delta$ the triangularity and the corresponding $s_X$ parameters
measure their shear.

\section{\label{sec:boundary} Insensitivity to boundary conditions}
Based on the size of the global term in the DKE \eqref{ldkes} -- which sets the
radial coupling -- the radial correlations are expected to decrease
outside the pedestal region. As a consequence, the flows and fluxes in
the pedestal are essentially decorrelated from the boundary
conditions, provided that the boundaries are sufficiently far away
from the pedestal.

\begin{figure}
  \includegraphics{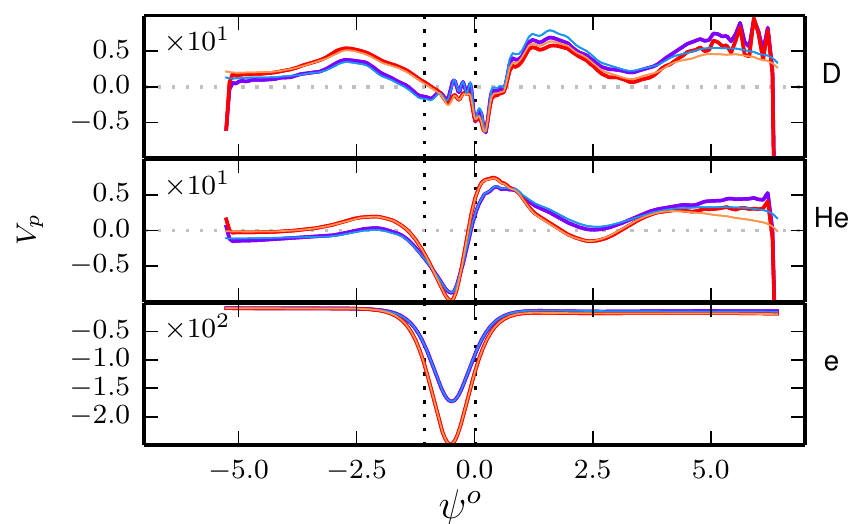}  
  \caption{\label{fig:neumann_flux_Vp} Global outboard particle flows in a fixed $\Phi$ scan with Dirichlet (thick) and Neumann (thin) boundary conditions.}
\end{figure}

To demonstrate this, we performed identical simulations with Neumann
boundary conditions ($\vec{v}_d \cdot \nabla g = 0$) instead of
Dirichlet ($g = g_\text{local}$). The resulting poloidal flows are
displayed in thin lines \autoref{fig:neumann_flux_Vp} and largely
overlap around the pedestal with the Dirichlet results indicated with
thick lines -- even though the Dirichlet boundary condition can introduce
massive oscillations near the boundaries.

The poloidal flows were chosen for this test as they are particularly
sensitive to numerical errors, since their evaluation involves a
derivative of simulation outputs [the final term in \eqref{eq:Vp}].
Similar or higher degrees of agreement are also found for the poloidal
variation of $V_p$ in the middle of the pedestal, and other quantities
such as the radial fluxes and sources.  Thus, we concluded that the
results indeed are largely independent of the boundary conditions.

\section{\label{sec:momentum} Radial current replaced by momentum sources}

We observe a finite radial current in our simulations, which is
balanced by non-quasineutral particle sources. If this would be the
only radial current, the corresponding time evolution of the electric
field would be inconsistent with the assumption of steady state. Even
though it cannot be ruled out that other, not modeled,
non-intrinsically ambipolar processes cancel our radial current, the
appearance of a radial current in the simulation may nevertheless be
concerning. Note however, that since we do not solve for the radial
electric field our current is not a source of any numerical
inconsistency in the simulation. 

The torque from the radial current appears in the species-summed
toroidal angular momentum equation, and in steady state it is balanced
by the divergence of the toroidal angular momentum flux, and momentum
sources, if any. In the simulations presented here no momentum sources
were considered, therefore it is possible that the current we
observe develops merely to balance the radial variation of the
momentum flux, and that a radial current may not be necessary if
we allow for momentum sources. Here we show results indicating that
this is the case, although a thorough investigation of the problem is
outside the scope of this paper.

Momentum sources can arise from numerous effects, a few candidates near the edge are:
the radial variation of turbulent
momentum fluxes, atomic physics processes \cite{omotani16}, or orbit
loss effects \cite{SD2012PRL}.
However, we will not dwell on the physical origin of the momentum sources,
but instead solve for the radial dependence of a momentum source profile by requiering the radial current to vanish,
$\sum_a Z_a \hat{\Gamma}_{a}=0$. This
requirement represents only one additional constraint per radial grid
point (in contrast to one for each species), and accordingly, it
allows us to solve for only one new radial profile. To eliminate the
corresponding degrees of freedom we assumed momentum sources for the
various species to be proportional to the mass and the concentration
of the species in the core, $\hat{S}_{ma} \propto m_a n_{0a}
\hat{S}_{m}$, and we solve only for $\hat{S}_{m}$.
The velocity
space structure of the momentum sources was taken so that they do not
contribute to parallel heat fluxes, and we assume the sources to be
poloidally uniform. These choices are not motivated by any particular
physical process, but should be sufficient for our purposes.

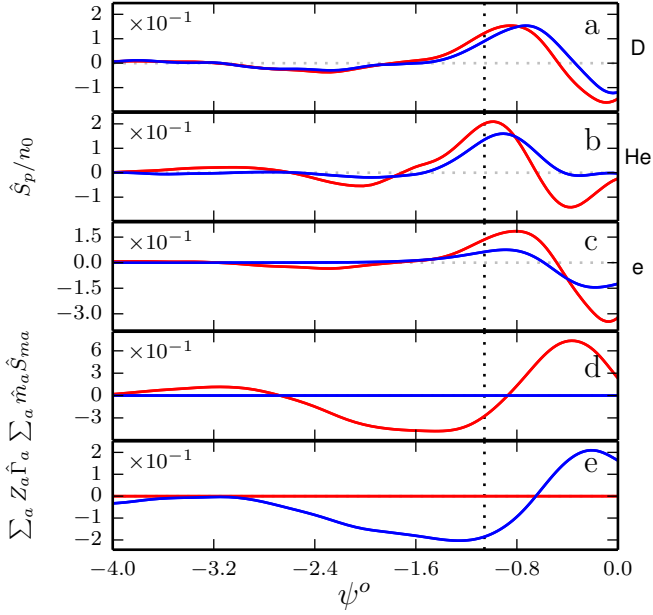
\begin{figure}
  \includegraphics{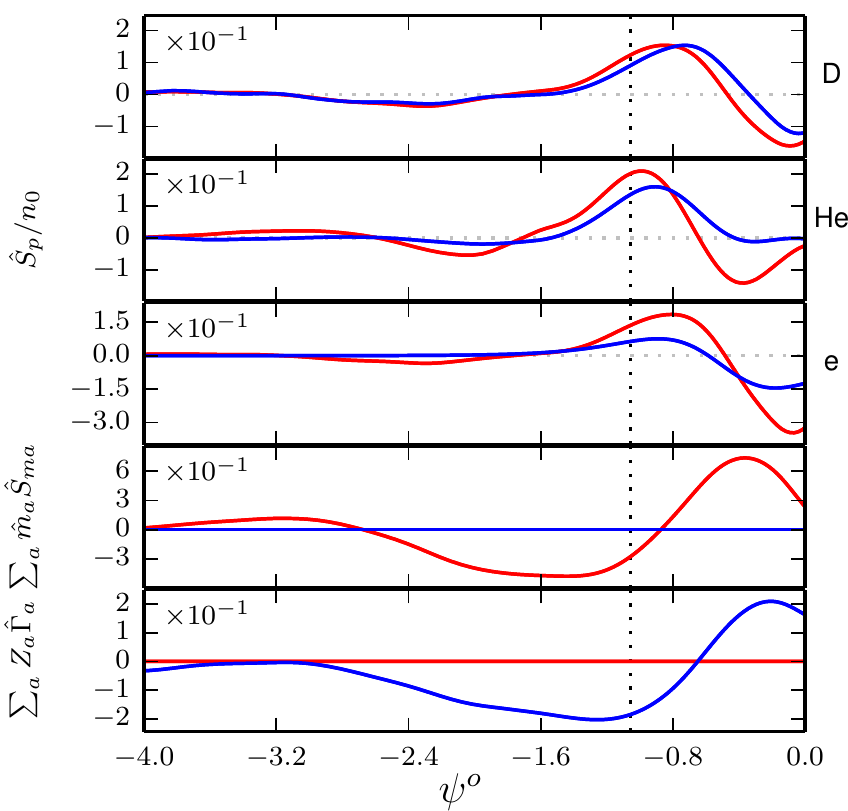}  
\put(-25,218){\large a}
\put(-25,174){\large b}
\put(-25,136){\large c}
\put(-25,88){\large d}
\put(-25,52){\large e}
  \caption{\label{fig:msource} Comparison of particle (a-c) and total
    momentum (d) sources, and radial current (e), between our baseline
    case (blue curves) and the case where the radial current is
    required to vanish (red curves).}
\end{figure}

Using the baseline deuterium bulk plasma case, we performed a
simulation where we allowed for momentum sources by the scheme specified above.
In \autoref{fig:msource} we show particle and momentum sources for the
simulation with zero radial current (red curves) and compare it to the
results without momentum sources (blue curves). As clearly seen in
\autoref{fig:msource}e, it is possible to enforce a vanishing radial
current (red curve is zero), but in that case there is a need for
finite momentum sources (see red curve in \autoref{fig:msource}d). The particle sources, shown in
\autoref{fig:msource}a-c, have changed to
achieve a zero net charge source. They still share
qualitative similarities with those in the simulation without momentum
sources: in both cases the particle sources have a positive peak a little inside the pedestal top and drop towards the separatrix in such a way that they
change sign around the middle of the model pedestal.

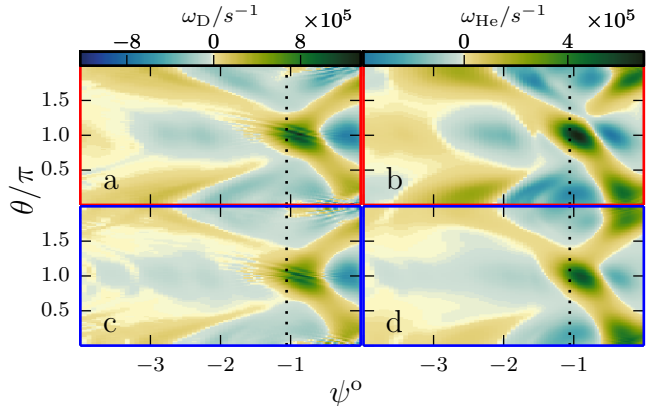
\begin{figure}
  \includegraphics{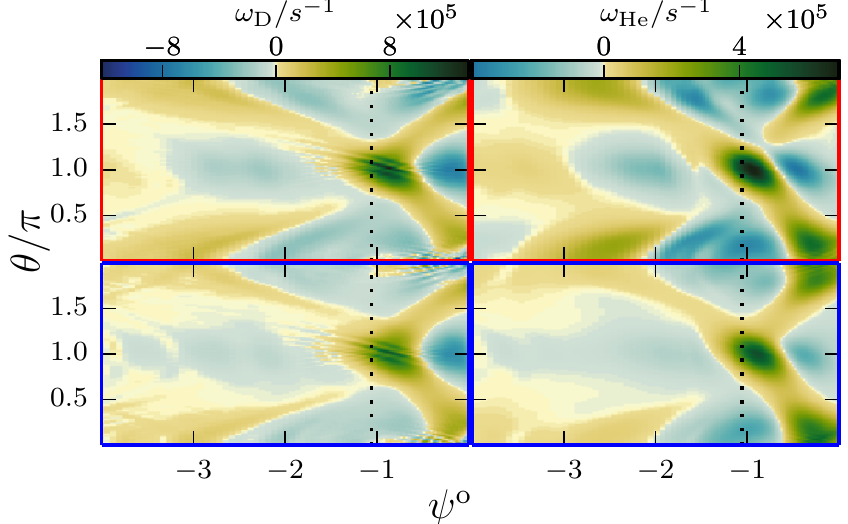} 
\put(-207,82){\large a}
\put(-100,82){\large b}
\put(-207,28){\large c}
\put(-100,28){\large d}  
  \caption{\label{fig:msource_vorticity} Vorticities of flows in the
    radial-poloidal plane for the baseline bulk deuterium plasma with
    (a-b) and without (c-d) momentum sources for the D (a,c) and He
    (b,d) species.}
\end{figure}

Importantly, the interaction of the radial and poloidal fluxes and the
corresponding complex flow patterns are not very sensitive to the
replacement of radial currents with momentum sources in the
simulation. This is perhaps best illustrated by comparing vorticities,
see \autoref{fig:msource_vorticity}, where (a-b) shows results with
zero radial current, and the case without momentum sources is shown in
(c-d).

While the issue deserves a more detailed study, here we have demonstrated
that radial currents are not a necessary feature of our simulations,
and that from the point of view of flows it is of secondary importance
whether the radial variation of the toroidal angular momentum is
balanced by a torque from a radial current or by momentum sources. The
red curve of \autoref{fig:msource}(d) and the blue curve of (e) are
proportional to the radial variation of the total toroidal angular
momentum flux in their respective cases ($\hat{S}_{ma}$ is here defined to make the proportionality constant the same in both cases).
Their similarity suggests
that the momentum fluxes are similar in the two cases. Without adding
a corresponding figure we note that this is indeed the case, although
the magnitude of the momentum transport is somewhat higher in the zero radial current case, as expected from comparing the magnitudes in \autoref{fig:msource}(d) and (e).

\bibliography{plasma-bib.bib} 
\end{document}